\documentclass[aps,prd,epsecnum,preprint,nofootinbib]{revtex4}
\usepackage{bm}
\usepackage{graphicx}
\usepackage{amsmath,amsthm,amssymb}

\begin{document}
\baselineskip5.5mm
\title{
Newtonian self-gravitating system in a relativistic huge void universe model 
}
\author{
       ${}^{1}$Ryusuke Nishikawa \footnote{E-mail:ryusuke@sci.osaka-cu.ac.jp},
	${}^{1}$Ken-ichi Nakao \footnote{E-mail:knakao@sci.osaka-cu.ac.jp},
and
       ${}^{2}$Chul-Moon Yoo \footnote{E-mail:yoo@gravity.phys.nagoya-u.ac.jp},
}
\affiliation{
${}^{1}$Department of Mathematics and Physics,
Graduate School of Science, Osaka City University,
3-3-138 Sugimoto, Sumiyoshi, Osaka 558-8585, Japan
\\
${}^{2}$Division of Particle and Astrophysical Science, 
Graduate School of Science, Nagoya University, 
Furo-cho, Chikusa-ku, Nagoya 464-8602, Japan
}

\begin{abstract}
\baselineskip5.5mm
We consider a test of the Copernican Principle through observations of the large-scale structures, 
and for this purpose we study the self-gravitating system 
in a relativistic huge void universe model which does not invoke the Copernican Principle. 
If we focus on the the weakly self-gravitating and slowly evolving system 
whose spatial extent is much smaller than the scale of the 
cosmological horizon in the homogeneous and isotropic background universe model,  
the cosmological Newtonian approximation is available. 
Also in the huge void universe model, the same kind of approximation as the cosmological 
Newtonian approximation is available for the analysis of the perturbations contained in 
a region whose spatial size is much smaller than the scale of the huge void:  the effects of the 
huge void are taken into account in a perturbative manner by using the Fermi-normal coordinates.  
By using this approximation, we derive the equations of motion 
for the weakly self-gravitating perturbations whose elements have relative velocities 
much smaller than the speed of light, 
and show the derived equations can be significantly different from 
those in the homogeneous and isotropic universe model,  due to the anisotropic volume expansion 
in the huge void.
We linearize the derived equations of motion and solve them.  
The solutions show that the behaviors of linear density perturbations are very 
different from those in the homogeneous and isotropic universe model.
\end{abstract}

\preprint{OCU-PHYS 410}

\preprint{AP-GR 115}

\maketitle

\section{introduction}\label{sec1}

Most of modern cosmological models are based on the Copernican principle 
which states the earth is not at a privileged position in the universe. 
The observed isotropy of the Cosmic Microwave Background (CMB) radiation 
together with the Copernican principle 
implies our universe is homogeneous and isotropic,
if the small scale structures less than 50 Mpc are coarse-grained.
Although the standard cosmology can explain a lot of observations naturally, 
we should note that the Copernican principle on the scale larger than 
1 Gpc has not been confirmed. 
This means modern cosmology would contain systematic errors
that arise from the inhomogeneities of the universe.
The systematic errors may mislead us 
when we consider major issues in modern cosmology such as 
the determination of cosmological parameters.
Thus, it is an unavoidable task in modern precision cosmology to test the Copernican principle. 

In order to test the Copernican principle, 
we have to investigate non-Copernican cosmological models which drop the Copernican principle. 
Non-Copernican models commonly assume that we live close to
the center in a spherically symmetric spacetime 
since the universe is observed to be nearly isotropic around us. 
These models have also been studied as an alternative to the model with the dark energy 
whose stress-energy tensor do not satisfy the strong energy condition so that all of the observational 
results until now are explained by the homogeneous and isotropic universe model 
in the framework of general relativity,
because some of them can explain the observation of Type Ia supernovae without introducing dark energy~\cite{Bull:2012zx,Celerier:1999hp,Celerier:2009sv,Clifton:2008hv,Goodwin:1999ej,Iguchi:2001sq,Kolb:2009hn,Mustapha:1998jb,Tomita:1999qn,Tomita:2000jj,Tomita:2001gh,Vanderveld:2006rb,Yoo:2008su,Yoo:2010qn}.
The non-Copernican models without dark energy have been tested by observations  
on the CMB acoustic peaks~\cite{Alexander:2007xx,Alnes:2005rw,Biswas:2010xm,Bolejko:2008cm,Clarkson:2010ej,GarciaBellido:2008nz,Marra:2011ct,Marra:2010pg,Moss:2010jx,Nadathur:2010zm,Yoo:2010qy,Zibin:2008vk},
the present Hubble parameter $H_0$~\cite{Biswas:2010xm,Bolejko:2008cm,GarciaBellido:2008nz,Marra:2011ct,Moss:2010jx},
the galaxy correlations on the Baryon Acoustic Oscillations (BAO) scale~\cite{Biswas:2010xm,GarciaBellido:2008yq,Zumalacarregui:2012pq},
the kinematic Sunyaev-Zeldovich (kSZ) effect~\cite{Bull:2011wi,GarciaBellido:2008gd,Moss:2011ze,Yoo:2010ad,Zhang:2010fa,Ade:2013opi}
and others~\cite{Adachi:2011vu,Alnes:2006uk,Bolejko:2011jc,Bolejko:2005fp,Caldwell:2013fua,Celerier:2012xr,Clarkson:2012bg,dePutter:2012zx,Dunsby:2010ts,Enqvist:2009hn,Enqvist:2006cg,Goto:2011ru,Heavens:2011mr,Quartin:2009xr,Regis:2010iq,Romano:2009mr,Romano:2010nc,Romano:2011mx,Tanimoto:2009mz,Tomita:2009ar,Yagi:2012vx,Zibin:2011ma},
and consequently significant observational constraints on these models exist. 
However, it should be noted that even if there are dark energy components, 
the existence of the large spherical inhomogeneity may significantly affects observational results
(see e.g. Ref.~\cite{Valkenburg:2013qwa,Valkenburg:2012td}). The huge void universe model which 
assumes we live near the symmetry center of the spherically symmetric void  
whose size is comparable to the radius of the cosmological horizon is known as 
the most popular non-Copernican model. In this paper, we consider 
the huge void universe model based on the Lema\^{\i}tre-Tolman-Bondi (LTB) solution 
which is an exact solution of the Einstein equations for the spherically symmetric 
spacetime filled with dust. 

Growth of the large-scale structures in the universe 
can be thought of as one of the 
most useful tools to examine the huge void universe model,
because the evolution of perturbations is expected to reflect the tidal force field 
in the background spacetime: The tidal force comes from the Weyl curvature, and hence  
there is the tidal force field, or simply, the tidal field in the huge void universe model but not in 
the homogeneous and isotropic universe model.
Recently, linear perturbations in the LTB cosmological model and the observations 
related to them have been studied by several researchers~\cite{Dunsby:2010ts,Alonso:2012ds,Alonso:2010zv,Clarkson:2009sc,February:2012fp,February:2013qza,Zibin:2008vj,Nishikawa:2012we,Nishikawa:2013rna,Nishikawa:2014sga}.

In our universe, there are well developed nonlinear structures,  
such as galaxies, clusters of galaxies and superclusters.
In the standard cosmology based on the homogeneous and isotropic universe model  
often called Friedmann-Lema\^{\i}tre-Robertson-Walker (FLRW) universe model, 
the dynamical evolutions of the perturbations corresponding to those structures are  
commonly studied by using the cosmological Newtonian approximation. 
The cosmological Newtonian approximation is applicable to the analysis of the dynamics 
of the perturbations which satisfy the following conditions (see, for example Ref.~\cite{peebles} and  
Refs.~\cite{Futamase:1989,Hwang:2005mg,Shibata:1995dg,Tomita:1988} for the Post-Newtonian extension); 
\begin{enumerate}
\item the length scale of the system is much smaller
than the radius of the cosmological horizon of the background universe model; 
\item the elements of the system have relative velocities much smaller than the speed of light and 
energy densities much larger than the stresses;  
\item the self-gravity of the system is not negligible but very weak.
\end{enumerate} 
Hereafter, we call the perturbations to which the cosmological Newtonian approximation is 
applicable {\it the cosmological Newtonian system or the cosmological Newtonian perturbations}. 

The equations of motion obtained by the cosmological Newtonian 
approximation for dark matter components 
are solved by the $N$-body simulation,
and their results have been compared with observational results of galaxy clustering.
The cosmological Newtonian systems have also been studied by some analytic approaches such as
the linear approximation and the Zel'dovich approximation, and 
these analyses have helped us to understand its gravitational instability. 
However, there is no practical approximation scheme to study 
the ``cosmological Newtonian system" in the huge void universe model, and hence 
we propose the one in this paper.

Although the huge void can be a non-linear structure and necessarily relativistic, 
a similar approximation as the cosmological Newtonian approximation is available to the perturbations 
in the huge void universe model, if the conditions similar to the three conditions for the 
validity of the cosmological Newtonian approximation are satisfied. 
However in the case of the huge void universe model, 
the first condition for the cosmological Newtonian approximation should be revised as  follows;  
\begin{enumerate}
\item the length scale of the system is much smaller than the {\it spacetime curvature radius} 
$\cal R$ of the background universe model.
\end{enumerate}
Note that $\cal R$ is not necessarily spatially constant and hence 
the original condition is a subset of the revised one. In the above condition, it is implicitly assumed 
that the length scale of the system is so small that $\cal R$ is almost spatially 
constant within the system. 
The tidal force produced by the void structure can be treated in a perturbative manner in the 
system that satisfies the revised condition 1, and such a 
perturbation scheme has been developed 
in studying weakly self-gravitating systems of the mass $m$ in the tidal field produced 
by a black hole with the mass much larger than $m$ by using the Fermi-normal 
coordinates~\cite{MM:1963,Ishii:2005xq}. Hereafter, 
following Ref.~\cite{Ishii:2005xq}, we call this approximation scheme {\it the tidal approximation}  
and will apply it to our problem. Hereafter, we call the system to which the tidal approximation is 
applicable simply {\it the Newtonian system or the Newtonian perturbations}. 
Of course, the Newtonian system implicitly 
corresponds to a galaxy, a cluster of galaxies or a supercluster, etc,  in the huge void.   

We denote the size and 
the typical velocity of the Newtonian system relative to the background by 
$\ell_{\rm N}$ and $\bm{v}_{\rm N}$, respectively.  
Then we introduce two non-negative small parameters defined as
\begin{eqnarray}
 \epsilon &:=&\frac{|\bm{v}_{\rm N}|}{c}, \label{5:twopara-e} \\
 \kappa&:=&\frac{\ell_{\rm N}}{\mathcal{R}},
 \label{5:twopara-k}
\end{eqnarray}
where $c$ is the speed of light. 
We note that both of the parameters $\epsilon$ and $\kappa$ are used as the expansion parameters of 
the tidal approximation.

This paper is organized as follows.
In \S~\ref{sec2}, after the brief review of the Fermi-normal coordinates,
we introduce the Fermi-normal coordinates in the huge void universe model. 
In \S~\ref{sec3}, we derive a set of equations governing the Newtonian system 
in the huge void universe model 
for three cases,
$\epsilon \gg \kappa$, $\epsilon\simeq \kappa$ and $\epsilon\ll\kappa$, individually.
In \S~\ref{sec4}, we solve the derived equations by using the linear approximation 
and investigate the growth of the vorticity field and the density perturbations.
\S~\ref{sec5} is devoted to conclusion and discussion.

In this paper, we use the geometrized unit in which both of the speed of light 
and the Newton's gravitational constant are one, but if necessary, we recover them:  
The speed of light and the Newton's gravitational constant are denoted by $c$ and $G$, respectively. 
The Latin indices denote the spatial components, whereas the Greek indices represent
the spacetime components.

\section{The LTB solution in Fermi-normal coordinates}\label{sec2}
\subsection{Definition of Fermi-normal coordinates}

First of all, we briefly review the Fermi-normal coordinates 
and the coordinate transformation from arbitrary coordinates to it 
(see, for detail, Refs.~\cite{Baldauf:2011bh,Klein:2007xj,toolkit,Schmidt:2012nw,Mashhoon:2007qm}). 
In this section, we denote the Fermi-normal coordinates by $x^\mu$ and the other by $x^{\mu'}$. 

Let $\gamma$ be a timelike geodesic; the components of its tangent vector 
with respect to the coordinate basis $\partial/\partial x^{\mu'}$ are denoted by
\begin{eqnarray}
 u^{\mu'}=\frac{dx^{\mu'}}{d\tau},
 \label{5:umu1}
\end{eqnarray}
where $\tau$ is the proper time measured along $\gamma$.
Then, we erect a parallelly transported orthonormal tetrad basis  
$e_{(\alpha)}^{\mu'}$ on $\gamma$:  
\begin{eqnarray}
 g_{\mu'\nu'}e_{(\alpha)}^{\mu'}e_{(\beta)}^{\nu'}=\eta_{(\alpha)(\beta)}~~{\rm and}~~
 u^{\mu'}\nabla_{\mu'}e_{(\alpha)}^{\nu'}=0,
 \label{5:tetrad1}
\end{eqnarray}
where $\eta_{(\alpha)(\beta)}={\rm diag}[-1,1,1,1]$,
and we assume $e_{(0)}^{\mu'}=u^{\mu'}$.  As usual, we denote 
the inverse matrix $\eta_{(\alpha)(\beta)}$ by $\eta^{(\alpha)(\beta)}$. Then, we define 
$e^{(\alpha)\mu'}=\eta^{(\alpha)(\beta)}e_{(\beta)}^{\mu'}$.  

In order to define the Fermi-normal coordinates which cover the 
neighborhood of $\gamma$, we focus on an event $P$ connected 
to $\gamma$ by an unique spacelike geodesic $\beta$ which orthogonally 
intersects $\gamma$.\footnote{If there is no such an unique spacelike geodesic, $P$ is not in the 
domain covered by the Fermi-normal coordinates associated to $\gamma$.} 
We call the intersection between $\gamma$ and $\beta$ the 
event $Q$. The components of the unit vector tangent to $\beta$ with 
respect to the coordinate basis $\partial/\partial x^{\mu'}$ are denoted by 
\begin{eqnarray}
 n^{\mu'}=\frac{dx^{\mu'}}{ds},
 \label{5:nmu1}
\end{eqnarray}
where $s$ is the proper length measured along $\beta$. 
We choose the origin of $s$ so that $s=0$ at $Q$. 
The components of the unit vector tangent to $\beta$ 
with respect to the tetrad basis at $s=0$ is given in the form 
\begin{equation}
e_{\mu'}^{(\alpha)} n^{\mu'}|_{s=0}=\left(0,\Omega^i\right)
 \label{5:Omega1}
 \end{equation}
by its definition. Note that $\Omega^i$ is normalized in the sense of $\delta_{ij}\Omega^i\Omega^j=1$. 
Conversely, $n^{\mu'}$ at $s=0$ is written as 
\begin{eqnarray}
 \left.n^{\mu'}\right|_{s=0}=\Omega^ie_{(i)}^{\mu'}.
 \label{5:Omega2}
\end{eqnarray}
We denote the proper time of $\gamma$ at the event $Q$ by $\tau$, 
whereas the proper length from $Q$ to $P$ along $\beta$ is denoted by $s$. 
Then, the values of the Fermi-normal coordinates at the event $P$ 
are defined as 
\begin{eqnarray}
 x^0=\tau~~{\rm and}~~x^i=s~\Omega^i.
 \label{5:fermi1}
\end{eqnarray}
The timelike geodesic $\gamma$ is called the fundamental timelike geodesic of this 
Fermi-normal coordinates.

In order to relate the original coordinates $x^{\mu'}$ 
to the Fermi-normal coordinates defined as Eq.~\eqref{5:fermi1}, 
we solve the geodesic equation to determine $\beta$ and obtain the 
solution in the form of the Maclaurin series as follows. The geodesic equation in the 
original coordinates $x^{\mu'}$ is 
\begin{eqnarray}
 \frac{d^2x^{\mu'}}{ds^2}
 +\Gamma^{\mu'}_{\alpha'\beta'}\frac{dx^{\alpha'}}{ds}\frac{dx^{\beta'}}{ds}=0.
 \label{5:geodesic1}
\end{eqnarray}
We write the solution $x^{\mu'}(s)$ and $\Gamma^{\mu'}_{\alpha'\beta'}\bigl(x^{\rho'}(s)\bigr)$ 
in the forms of the Maclaurin series, respectively, as
\begin{eqnarray}
 x^{\mu'}(s)&=&\sum_{N=0}\frac{s^N}{N!}\left.\left(\frac{d}{ds^N}x^{\mu'}\right)\right|_{s=0}
 =:\sum_{N=0}\frac{s^N}{N!}x^{\mu'}_{(N)}, 
 \label{5:expand1} \\
 \Gamma^{\mu'}_{\alpha'\beta'}\bigl(x^{\rho'}(s)\bigr)&=&\sum_{N=0}\frac{s^N}{N!}
 \left.\left(\frac{d}{ds^N}\Gamma^{\mu'}_{\alpha'\beta'}\right)\right|_{s=0}.
 \label{5:expand2}
\end{eqnarray}
From Eq.~\eqref{5:Omega2}, we have
\begin{eqnarray}
 x^{\mu'}_{(1)}=\Omega^ie_{(i)}^{\mu'}.
 \label{5:s1}
\end{eqnarray}
By substituting Eqs.~\eqref{5:expand1} and \eqref{5:expand2} into Eq.~\eqref{5:geodesic1}
and by using Eq.~\eqref{5:s1}, we obtain 
\begin{eqnarray}
 x^{\mu'}_{(2)}
 &=&
 -\left.\Gamma^{\mu'}_{\alpha'\beta'}\right|_{s=0}
 e^{\alpha'}_{(i)}e^{\beta'}_{(j)}\Omega^i\Omega^j,
 \label{5:s2} \\
 x^{\mu'}_{(3)}
 &=&
 \left(
 2\left.\Gamma^{\mu'}_{\alpha'\beta'}\right|_{s=0}
 \left.\Gamma^{\beta'}_{\gamma'\delta'}\right|_{s=0}
 -\left.\partial_{\delta'}\Gamma^{\mu'}_{\alpha'\gamma'}\right|_{s=0}
 \right)
 e^{\alpha'}_{(i)}e^{\gamma'}_{(j)}e^{\delta'}_{(k)}\Omega^i\Omega^j\Omega^k.
 \label{5:s3}
\end{eqnarray}
By substituting Eqs.~\eqref{5:s1}--\eqref{5:s3} into Eq.~\eqref{5:expand1} and by using Eq.~\eqref{5:fermi1},
we obtain
\begin{eqnarray}
 x^{\mu'}&=&x^{\mu'}_{(0)}+e_{(i)}^{\mu'}x^i
 -\frac{1}{2}\left.\Gamma^{\mu'}_{\alpha'\beta'}\right|_{s=0}
 e^{\alpha'}_{(i)}e^{\beta'}_{(j)}x^ix^j \cr
 & & \cr
 &+&\frac{1}{6}\left(
 2\left.\Gamma^{\mu'}_{\alpha'\beta'}\right|_{s=0}\left.\Gamma^{\beta'}_{\gamma'\delta'}\right|_{s=0}
 -\left.\partial_{\delta'}\Gamma^{\mu'}_{\alpha'\gamma'}\right|_{s=0}
 \right)
 e^{\alpha'}_{(i)}e^{\gamma'}_{(j)}e^{\delta'}_{(k)}x^ix^jx^k~
 +\mathcal{O}\left(|\bm{x}|^4\right), 
 \label{5:trans1}
\end{eqnarray}
where 
$$
|\bm{x}|^2:=\delta_{ij}x^i x^j.
$$
By differentiating Eq.~\eqref{5:trans1} with respect to $x^0$, we obtain
\begin{eqnarray}
 \frac{\partial x^{\mu'}}{\partial x^0}&=&
 \frac{\partial x^{\mu'}_{(0)}}{\partial x^0}
 +x^i\frac{\partial}{\partial x^0}e_{(i)}^{\mu'}
 -\frac{1}{2}x^ix^j
 \frac{\partial}{\partial x^0}\left(
 \left.\Gamma^{\mu'}_{\alpha'\beta'}\right|_{s=0}e^{\alpha'}_{(i)}e^{\beta'}_{(j)}
 \right)~+~\mathcal{O}\left(|\bm{x}|^3\right),
 \cr
 &&\cr
 &=&
 e_{(0)}^{\mu'}-\left.\Gamma^{\mu'}_{\alpha'\beta'}\right|_{s=0}e_{(0)}^{\alpha'}e_{(j)}^{\beta'}x^j
 -\left.\frac{1}{2}
 \left(\partial_{\delta'}\Gamma^{\mu'}_{\alpha'\beta'}
 -2\Gamma^{\mu'}_{\alpha'\tau'}\Gamma^{\tau'}_{\delta'\beta'}\right)
 \right|_{s=0}
 e_{(0)}^{\delta'}e_{(j)}^{\alpha'}e_{(k)}^{\beta'}x^jx^k \cr
 &&\cr
 &&~+~\mathcal{O}\left(|\bm{x}|^3\right),
 \label{5:trans2}
\end{eqnarray}
where in the second equality we have used the relations
\begin{eqnarray}
 \frac{\partial}{\partial x^0}e^{\nu'}_{(j)}\Bigr|_{s=0}
 =-\left.\Gamma^{\nu'}_{\alpha'\beta'}\right|_{s=0}e_{(0)}^{\alpha'}e_{(j)}^{\beta'}~~{\rm and}~~
 \frac{\partial}{\partial x^0}
 \left.\Gamma^{\mu'}_{\alpha'\beta'}\right|_{s=0}=e_{(0)}^{\delta'}
 \left.\partial_{\delta'}\Gamma^{\mu'}_{\alpha'\beta'}\right|_{s=0}.
 \label{5:relation1}
\end{eqnarray}
By differentiating Eq.~\eqref{5:trans1} with respect to $x^i$, we obtain
\begin{eqnarray}
 \frac{\partial x^{\mu'}}{\partial x^i}&=&
 e_{(i)}^{\mu'}
 -\left.\Gamma^{\mu'}_{\alpha'\beta'}\right|_{s=0}
 e^{\alpha'}_{(i)}e^{\beta'}_{(j)}x^j \cr
 &&\cr
 &&+\frac{1}{6}\left.\left(
 4\Gamma^{\mu'}_{\alpha'\tau'}\Gamma^{\tau'}_{\delta'\beta'}
 +2\Gamma^{\mu'}_{\delta'\tau'}\Gamma^{\tau'}_{\alpha'\beta'}
 -2\partial_{\alpha'}\Gamma^{\mu'}_{\delta'\beta'}
 -\partial_{\delta'}\Gamma^{\mu'}_{\alpha'\beta'}
 \right)\right|_{s=0}e_{(i)}^{\delta'}e_{(j)}^{\alpha'}e_{(k)}^{\beta'}x^j x^k \cr
 &&\cr
 &&+\mathcal{O}(|\bm{x}|^3).
 \label{5:trans3}
\end{eqnarray}
Eqs.~\eqref{5:trans2} and \eqref{5:trans3} are written in the following unified form;
\begin{eqnarray}
\frac{\partial x^{\mu'}}{\partial x^\nu}&=&e^{\alpha'}_{(\nu)}\biggl[\delta^{\mu'}_{\alpha'}
 -\left.\Gamma^{\mu'}_{\alpha'\beta'}\right|_{s=0}e_{(j)}^{\beta'}x^j
 -\left.\frac{1}{2}
 \left(\partial_{\alpha'}\Gamma^{\mu'}_{\beta'\gamma'}
 -2\Gamma^{\mu'}_{\beta'\delta'}\Gamma^{\delta'}_{\alpha'\gamma'}\right)
 \right|_{s=0}
e_{(j)}^{\beta'}e_{(k)}^{\gamma'}x^jx^k \cr
 &&\cr
 &-&\frac{1}{3}\left(\delta_{\alpha'}^{\delta'}+e_{(0)}^{\delta'}e_{(0)\alpha'}\right)
 R_{\delta' \beta' \gamma'}{}^{\mu'}
 e_{(j)}^{\beta'}e_{(k)}^{\gamma'}x^jx^k
\biggr]
+\mathcal{O}\left(|\bm{x}|^3\right).
\label{5:trans_unif}
\end{eqnarray}
Eq.~\eqref{5:trans_unif} is the coordinate transformation matrix for the covariant components 
of any tensors from the original coordinates $x^{\mu'}$ to the Fermi-normal coordinates $x^\nu$.

\subsection{Components of stress-energy and metric tensors in Fermi-normal coordinates}

The components of the stress-energy tensor with respect to the original coordinate basis is written as
\begin{equation}
T_{\mu'\nu'}=\rho' u_{\mu'} u_{\nu'},  \label{5:tmunu1}
\end{equation}
where $\rho'(x^{\mu'})$ and $u^{\mu'}(x^{\nu'})$ are the energy density and the 4-velocity field 
of the dust, respectively.

We compute the components of the 4-velocity field of the dust with respect to 
the Fermi-normal coordinates by using the coordinate transformation~\eqref{5:trans_unif} first.
The covariant components of the 4-velocity $u_\mu$ in the Fermi-normal coordinates are given as 
\begin{eqnarray}
 u_\mu (x^\rho) =\frac{\partial x^{\nu'}}{\partial x^\mu}u_{\nu'}(x^{\rho'}).
 \label{5:umu2}
\end{eqnarray}
The covariant components $u_{\nu'}$ with respect to the original coordinate basis 
is written in the form of the Maclaurin series around the fundamental timelike geodesic 
$\gamma$, i.e., $s=0$ as 
\begin{eqnarray}
 u_{\nu'}(x^{\rho'})=\left.u_{\nu'}\right|_{s=0}+\left.\partial_{\alpha'}u_{\nu'}\right|_{s=0}\delta x^{\alpha'}
 +\frac{1}{2}\left.\partial_{\alpha'}\partial_{\beta'}u_{\nu'}\right|_{s=0}\delta x^{\alpha'}\delta x^{\beta'}
 +\cdot\cdot\cdot,
 \label{5:umu3}
\end{eqnarray}
where $\delta x^{\alpha'}$ is defined as 
\begin{eqnarray}
 \delta x^{\alpha'}:=x^{\alpha'}-\left.x^{\alpha'}\right|_{s=0}=e_{(j)}^{\alpha'}x^j
 -\frac{1}{2}\Gamma^{\alpha'}_{\tau'\rho'}e_{(j)}^{\tau'}e_{(k)}^{\rho'}x^jx^k
 +\mathcal{O}\left(|\bm{x}|^3\right),
 \label{5:deltax}
\end{eqnarray}
where we have used Eq.~\eqref{5:trans1} in the second equality. 
By substituting Eqs.~\eqref{5:trans_unif}, \eqref{5:umu3} and \eqref{5:deltax}
into Eq.~\eqref{5:umu2}, we obtain 
\begin{eqnarray}
 u_\mu (x^{\rho})&=&\left.u_{\alpha'}\right|_{s=0}e_{(\mu)}^{\alpha'}
 +\left.\left(
 \partial_{\beta'}u_{\alpha'}-\Gamma^{\tau'}_{\alpha'\beta'}u_{\tau'}
 \right)\right|_{s=0}
 e^{\alpha'}_{(\mu)}e^{\beta'}_{(j)}x^j
 +\mathcal{O}\left(|\bm{x}|^2\right).
 \label{5:umu4}
\end{eqnarray}

The energy density $\rho$ in the Fermi-normal coordinates is given by
\begin{eqnarray}
\rho(x^\mu) &=& \rho'(x^{\mu'}) \cr
&&\cr
&=&\left.\rho'(x^{\mu'})\right|_{s=0}
 +\left.\partial_{\alpha'}\rho'(x^{\mu'})\right|_{s=0}\delta x^{\alpha'}
 +\cdots \cr
 && \cr
 &=&\left.\rho'(x^{\mu'})\right|_{s=0}
 +\left.\partial_{\alpha'}\rho'(x^{\mu'})\right|_{s=0}e^{\alpha'}_{(j)}x^j
 +\mathcal{O}\left(|\bm{x}|^2\right),
 \label{5:rho1}
\end{eqnarray}
where we have used Eq.~\eqref{5:deltax} in the second equality.

After lengthy but straightforward calculations, we obtain the metric in the Fermi-normal 
coordinates as 
\begin{subequations}
\begin{align}
 g_{00}&= -1-\hat{R}_{0i0j}x^ix^j+\mathcal{O}\left(|\bm{x}|^3\right),
 \label{5:met1a} \\
 g_{0i}&= -\frac{2}{3}\hat{R}_{0jik}x^jx^k
 +\mathcal{O}\left(|\bm{x}|^3\right),
 \label{5:met1b} \\
 g_{ij}&= \delta_{ij}-\frac{1}{3}\hat{R}_{ikjl}x^kx^l
 +\mathcal{O}\left(|\bm{x}|^3\right),
 \label{5:met1c}
\end{align}
\end{subequations}
where $\hat{R}_{\mu \nu \rho \sigma}$ is defined as
\begin{eqnarray}
 \hat{R}_{\mu \nu \rho \sigma}(x^0)
 &:=&e^{\alpha'}_{(\mu)}e^{\beta'}_{(\nu)}e^{\gamma'}_{(\rho)}e^{\delta'}_{(\sigma)}
 \left.R_{\alpha' \beta' \gamma' \delta'}\right|_{s=0},
 \label{5:Riemann1}
\end{eqnarray}
where $R_{\alpha' \beta' \gamma' \delta'}$ represents the components of the 
Riemann tensor with respect to the original coordinate basis $\partial/\partial x^{\mu'}$. 
For later convenience, we define $h^{\rm B}_{\mu \nu}$ as
\begin{eqnarray}
 h^{\rm B}_{00}:=-\hat{R}_{0i0j}x^ix^j,~~~~
 h^{\rm B}_{0i}:=-\frac{2}{3}\hat{R}_{0jik}x^jx^k,~~~~
 h^{\rm B}_{ij}:=-\frac{1}{3}\hat{R}_{ikjl}x^kx^l.
 \label{5:Hmunu1}
\end{eqnarray}

As mentioned in \S~\ref{sec1}, we consider the perturbations 
of the length scale $\ell_{\rm N}\ll {\cal R}$, and hence 
we have introduced a small parameter $\kappa$ by Eq.~\eqref{5:twopara-k}. 
If we analyze the behaviors of such perturbations in the Fermi-normal coordinates, the condition 
$|\bm{x}|=\mathcal{O}(\ell_{\rm N}) \ll {\cal R}$  is always satisfied in the domain of our interest. 
Hence we have
\begin{eqnarray}
h^{\rm B}_{\mu\nu}
 =\mathcal{O}\left(\frac{|\bm{x}|^2}{\mathcal{R}^2}\right)
 =\mathcal{O}(\kappa^2).
 \label{5:kappa1}
\end{eqnarray}
Then, by adopting $\kappa$ as a book-keeping parameter which will be taken out of equations  
after counting the order of magnitude, and by using Eqs.~\eqref{5:Hmunu1} and
\eqref{5:kappa1}, we rewrite the components of the metric tensor given in Eqs.~\eqref{5:met1a}--\eqref{5:met1c}  as
\begin{eqnarray}
 g_{\mu \nu}&=&\eta_{\mu \nu}+h^{\rm B}_{\mu \nu}+\mathcal{O}(\kappa^3).
 \label{5:met2}
\end{eqnarray}
If we will analyze the only leading order effects, we can  
ignore the higher order terms with respect to $\kappa$.

\subsection{The Fermi-normal coordinates in the huge void universe model }

Here, we perform the coordinate transformation from 
the original coordinates of the huge void universe model based on the LTB solution 
to the Fermi-normal ones. The line element of the LTB solution is given 
in the synchronous comoving coordinates as follows;
\begin{equation}
 ds^2=
 -dt'^2+\frac{[\partial_{r'}R(t',r')]^2}{1-k(r')}dr'^2+R^2(t',r')
 \left(d\theta'^2+\sin^2\theta' d\phi'^2\right),
 \label{5:met3}
 \end{equation}
where we have denoted the original coordinates by $(x^{0'},x^{1'},x^{2'},x^{3'})=(t',r',\theta',\phi')$.
For later convenience, we define Hubble functions as 
\begin{eqnarray}
H_\parallel (t',r')&:=&\frac{\partial_{t'}\partial_{r'}R(t',r')}{\partial_{r'}R(t',r')}, \\
&&\cr
H_\bot (t',r')&:=&\frac{\partial_{t'}R(t',r')}{R(t',r')}. 
\end{eqnarray}
As mentioned, the spacetime is filled with dust whose stress-energy tensor is given  
by \eqref{5:tmunu1}. 
The original synchronous and comoving coordinates are chosen so that   
the components of the 4-velocity field is given by $u^{\mu'}=\delta_{0'}^{\mu'}$.  
Each fluid element of the dust moves along a timelike geodesic whose unit tangent vector 
agrees with $u^{\mu'}$.

We choose a world line of a fluid element of the dust which stays at
constant spatial coordinates $(r',\theta',\phi')=(r'_{\rm o},\theta'_{\rm o},\phi'_{\rm o})$
as the fundamental timelike geodesic $\gamma$.  
It should be noted that 
since the original time coordinate $t'$ agrees with the proper time of the fundamental timelike 
geodesic $\gamma$, 
$t'$ along $\gamma$ agrees 
with the time coordinate $x^0$ of the Fermi-normal coordinate system. 
Then, the following parallelly transported tetrad basis is convenient for our purpose; 
\begin{subequations}
\begin{align}
 e_{(0)}^{\mu'}&=(1,0,0,0),
 \label{5:tet1a} \\
 e_{(1)}^{\mu'}&=\left(0,\sqrt{1-k(r'_{\rm o})}/\partial_{r'}R(t',r'_{\rm o}),0,0\right),
 \label{5:tet1b} \\
 e_{(2)}^{\mu'}&=\left(0,0,1/R(t',r'_{\rm o}),0\right),
 \label{5:tet1c} \\
 e_{(3)}^{\mu'}&=\left(0,0,0,1/R(t',r'_{\rm o})\sin \theta'_{\rm o}\right).
 \label{5:tet1d}
\end{align}
\end{subequations}

In the previous subsection, 
we have obtained the energy density and the components of 4-velocity field in the Fermi-normal 
coordinates as Eqs.~\eqref{5:umu4} and \eqref{5:rho1} in the form of the Maclaurin series with respect to 
the spatial coordinates $x^i$. 
Here, we should note that the term of the higher power in this series corresponds to the 
higher order term with respect to the parameter $\kappa$ defined as Eq.~\eqref{5:twopara-k} 
in the case of the huge void universe model. The size of the void is the same order 
as the cosmological horizon scale. Since the cosmological horizon scale is the same order as the 
spacetime curvature radius $\cal R$, and hence the $n$-th spatial derivatives of $u^{\mu'}$ and $\rho'$ 
are the same orders as themselves divided by ${\cal R}^n$, respectively. 
Since, as mentioned, $|\bm{x}|/{\cal R}={\cal O}(\kappa)$, we have
\begin{equation}
(x^i\partial_i)^n u^{\mu'} ={\cal O}(\kappa^n u^{\mu'})~~~~{\rm and}~~~~
(x^i\partial_i)^n \rho'={\cal O}(\kappa^n  \rho').
\end{equation}

By substituting Eqs.~\eqref{5:tmunu1} and \eqref{5:tet1a}--\eqref{5:tet1d} into Eq.~\eqref{5:rho1},
the energy density $\rho$ in the Fermi-normal coordinates is given by
\begin{eqnarray}
 \rho (x^\mu) &=&\rho_{\rm B}(x^0)
 +\left.\left[
 \left(\frac{\sqrt{1-k(r')}}{\partial_{r'}R(t',r')}\right)\partial_{r'}\rho' (t',r')
 \right]
 \right|_{t'=x^0,r'=r'_{\rm o}}x^1
 +\mathcal{O}\left(\kappa^2\rho_{\rm B}\right),
 \label{5:rho2}
\end{eqnarray}
where 
\begin{equation}
\rho_{\rm B}(x^0):=\left.\rho'(t',r')\right|_{t'=x^0,r'=r'_{\rm o}}.
\end{equation}

By substituting Eqs.~\eqref{5:tmunu1} and \eqref{5:tet1a}--\eqref{5:tet1d} into Eq.~\eqref{5:umu4},
we obtain the components of the 4-velocity field in the Fermi-normal coordinates as 
\begin{subequations}
 \begin{align}
 u^0(x^\mu)&=1~+\mathcal{O}(\kappa^2), \label{5:umu5a} \\
 u^1(x^\mu)&=H_\parallel^{\rm B}(x^0)x^1+\mathcal{O}(\kappa^2),
 \label{5:umu5b} \\
 u^2(x^\mu)&=H_\bot^{\rm B}(x^0)x^2+\mathcal{O}(\kappa^2),
 \label{5:umu5c} \\
 u^3(x^\mu)&=H_\bot^{\rm B}(x^0)x^3+\mathcal{O}(\kappa^2),
 \label{5:umu5d}
 \end{align}
\end{subequations}
where we have defined two kinds of local Hubble functions as
\begin{eqnarray}
H^{\rm B}_\parallel (x^0)&:=& \left. H_\parallel (t',r')\right|_{t'=x^0,r'=r'_{\rm o}}, \\
H^{\rm B}_\bot (x^0)       &:=& \left.H_\bot (t',r')\right|_{t'=x^0,r'=r'_{\rm o}}.
\end{eqnarray}
For later discussion, we define the 3-velocity as 
\begin{equation}
v^i:=\frac{u^i}{u^0}. \label{3-veloity-def}
\end{equation}
By using Eqs.~\eqref{5:umu5a}--\eqref{5:umu5d},
the 3-velocity $v^i$ in the Fermi-normal coordinates is given by
\begin{eqnarray}
 v^i=H^i{}_j(x^0)x^j+\mathcal{O}(\kappa^2),
 \label{5:velocity1}
\end{eqnarray}
where we have introduced $H^i{}_j$ defined as
\begin{eqnarray}
 H^i{}_j(x^0):=
 \left(
 \begin{array}{ccc}
 H^{\rm B}_\parallel (x^0) & 0 & 0 \\
 0 & H^{\rm B}_\bot (x^0) & 0 \\
 0 & 0 & H^{\rm B}_\bot (x^0) \\
 \end{array}
 \right).
 \label{5:matrix1}
\end{eqnarray}

By substituting Eqs.~\eqref{5:met3} and \eqref{5:tet1a}--\eqref{5:tet1d} into Eq.~\eqref{5:Hmunu1}
and by computing Riemann tensors in the original coordinates,
we obtain $h^{\rm B}_{\mu\nu}$ as
\begin{subequations}
\begin{align}
 h^{\rm B}_{00}&= -K_1(x^0)(x^1)^2-K_2(x^0)\left[(x^2)^2+(x^3)^2\right], \label{5:H00} \\
 h^{\rm B}_{01}&=0=h^{\rm B}_{02}= h^{\rm B}_{03}~, \\ 
 h^{\rm B}_{11}&= -\frac{1}{3}K_3(x^0)\left[(x^2)^2+(x^3)^2\right], \\
 h^{\rm B}_{12}&= \frac{1}{3}K_3(x^0)x^1x^2, \\
 h^{\rm B}_{13}&= \frac{1}{3}K_3(x^0)x^1x^3, \\
 h^{\rm B}_{22}&= -\frac{1}{3}K_3(x^0)(x^1)^2-\frac{1}{3}K_4(x^0)(x^3)^2, \\
 h^{\rm B}_{23}&= \frac{1}{3}K_4(x^0)x^2x^3, \\
 h^{\rm B}_{33}&= -\frac{1}{3}K_3(x^0)(x^1)^2-\frac{1}{3}K_4(x^0)(x^2)^2,
\end{align}
\end{subequations}
where $K_1(x^0)$, $K_2(x^0)$, $K_3(x^0)$ and $K_4(x^0)$ are defined as
\begin{eqnarray}
 K_1(x^0)&=& -\left.\frac{\partial_{t'}^2\partial_{r'}R(t',r')}{\partial_{r'}R(t',r')}\right|_{t'=x^0,r'=r'_{\rm o}}, \\
 K_2(x^0)&=& -\left.\frac{\partial_{t'}^2R(t',r')}{R(t',r')}\right|_{t'=x^0,r'=r'_{\rm o}}, \\
 K_3(x^0)&=& \left.\left[H_{\parallel}(t',r')H_\bot(t',r')
 +\frac{\partial_{r'}k(r')}{2R(t',r')\partial_{'r}R(t',r')}\right]\right|_{t'=x^0,r'=r'_{\rm o}}, \\
 K_4(x^0)&=& \left.\left[H^2_\bot(t',r')+\frac{k(r')}{R^2(t',r')}\right]\right|_{t'=x^0,r'=r'_{\rm o}}.
\end{eqnarray}

From Eq.~\eqref{5:trans1},
the original coordinates are related to the Fermi-normal coordinates by 
\begin{eqnarray}
 t'&=& x^0- \frac{\kappa}{2}\left[(x^1)^2H^{\rm B}_\parallel (x^0)
 +\left\{(x^2)^2+(x^3)^2\right\}H^{\rm B}_\bot (x^0)\right]
 +\mathcal{O}\left(\kappa^2|\bm{x}|\right) ,\\
 r'-r'_{\rm o}&=& \frac{\sqrt{1-k(r'_{\rm o})}}{\left.\partial_{r'}R(x^0,r')\right|_{r'=r'_0}}~x^1
 +\mathcal{O}\left(\kappa|\bm{x}|\right), \\
 \theta' -\theta'_{\rm o}&=& \frac{1}{R(x^0,r'_{\rm o})}~x^2
 +\mathcal{O}\left(\kappa\right) ,\\
 \phi' -\phi'_{\rm o}&=& \frac{1}{R(x^0,r'_{\rm o})\sin\theta'_{\rm o}}~x^3
 +\mathcal{O}\left(\kappa\right) .
\end{eqnarray}

We derive the basic equations up to the leading order
with respect to $\kappa$ for the LTB solution in the Fermi-normal coordinates. For this purpose, 
we denote the leading order of the 3-velocity $v^i$ by $v^i_{\rm B}$, i.e., 
\begin{equation}
 v^i_{\rm B}(x^\mu):=H^i{}_j(x^0)x^j, 
 \label{5:local1-v}
\end{equation}
 and define the ``gravitational potential" $\Phi_{\rm B}$ as 
\begin{equation}
\Phi_{\rm B}(x^\mu):=-\frac{1}{2}h^{\rm B}_{00}(x^\mu).
 \label{5:local1-phi}
\end{equation}
The Einstein equations for the LTB solution lead to
\begin{eqnarray}
 \frac{1}{3}\frac{\partial_{t'}^2\partial_{r'}R(t',r')}{\partial_{r'}R(t',r')}
 +\frac{2}{3}\frac{\partial_{t'}^2R(t',r')}{R(t',r')}
 &=&-\frac{4\pi}{3}\rho(t',r'),
 \label{2:poisson} \\
 && \cr
 \partial_{t'}\rho (t',r')+\left[H_\parallel (t',r')+2H_\bot (t',r')\right]\rho (t',r')
 &=&0.
 \label{2:conservation}
\end{eqnarray}
Then from Eq.~\eqref{2:conservation}, we have the leading order of the energy conservation law 
in the form
\begin{eqnarray}
 \partial_0\rho_{\rm B}+\partial_j
 \left(\rho_{\rm B}v^j_{\rm B}\right)&=& 0.
 \label{5:cons1}
\end{eqnarray}
By using Eqs.~\eqref{5:matrix1} and \eqref{5:H00}, we obtain the relation between 
$v_{\rm B}^i$ and $\Phi_{\rm B}$, which corresponds to the Euler equations, as
\begin{eqnarray}
 \partial_0v_{\rm B}^i+v_{\rm B}^j\partial_jv_{\rm B}^i
 &=& -\partial_i\Phi_{\rm B}.
 \label{5:euler1}
\end{eqnarray}
By using Eq.~\eqref{2:poisson}, we obtain the relation between 
$\rho_{\rm B}$ and $\Phi_{\rm B}$, which corresponds to  
the Poisson equation for the gravitational potential, as
\begin{eqnarray}
 \nabla^2\Phi_{\rm B}&=& 4\pi \rho_{\rm B},
 \label{5:poisson1}
\end{eqnarray}
where $\nabla^2=\delta^{ij}\partial^2/\partial x^i\partial x^j$.
By definition of $\Phi_{\rm B}$, we have 
$\nabla^2\Phi_{\rm B}=\mathcal{O}(\mathcal{R}^{-2})$.
Hence, from Eq.~\eqref{5:poisson1}, we have
\begin{equation}
\rho_{\rm B}=\mathcal{O}(\mathcal{R}^{-2}). \label{rhoB-order}
\end{equation}

From Eqs.~\eqref{5:cons1}--\eqref{5:poisson1},
we can see that the leading order of the basic equations for the LTB solution 
in the Fermi-normal coordinates take  the same forms as 
those of equations for the Newtonian self-gravitating system of the homogeneous dust fluid.

\section{Basic equations for the perturbations in the huge void universe 
model by the tidal approximation}\label{sec3}
In this section, we derive the basic equations governing the 
Newtonian perturbations in the huge void universe model 
based on the LTB solution in the framework of the tidal approximation. 
Hereafter, we assume that the background huge void universe model is covered 
by the Fermi-normal coordinates and denote the background quantities by the symbols with an over-bar. 
The components of the metric tensor of the huge void universe model with the Newtonian 
perturbations is written in the form
\begin{equation}
 g_{\mu\nu}=\bar{g}_{\mu\nu}+h^{\rm N}_{\mu\nu}. \label{5:schematic1}
\end{equation}
The energy density and the components of the 4-velocity field of the dust are written as
\begin{eqnarray}
\rho&=&\bar{\rho}+\rho_{\rm N}, \\
u^\mu&=&\bar{u}^\mu+\delta u^\mu.
\end{eqnarray}
Then the stress-energy tensor of the dust is 
\begin{equation}
T^{\mu\nu}=\bar{T}^{\mu\nu}+\delta T^{\mu\nu},\label{5:schematic2}
\end{equation}
where
\begin{eqnarray}
\bar{T}^{\mu\nu}&=&\bar{\rho}\bar{u}^\mu\bar{u}^\nu, \\
\delta T^{\mu\nu}&=&\bar{\rho}\left(2\bar{u}^{(\mu}\delta u^{\nu)}+\delta u^\mu \delta u^\nu\right)
+\rho_{\rm N} (\bar{u}^\mu+\delta u^\mu) (\bar{u}^\nu+\delta u^\nu).
\end{eqnarray}
As shown in the previous section, 
the components of the metric and stress-energy tensors of the background up to the 
non-trivial leading order are given by 
\begin{eqnarray}
 \bar{g}_{\mu\nu}&=&\eta_{\mu\nu}+h^{\rm B}_{\mu\nu}, \label{5:metricB} \\
 \bar{T}^{00}&=&\rho_{\rm B}=\mathcal{O}(\mathcal{R}^{-2}), \label{5:t00B} \\
 \bar{T}^{0i}&=&\rho_{\rm B}v^i_{\rm B}=\mathcal{O}(\kappa\mathcal{R}^{-2}), 
 \label{5:t0iB} \\
 \bar{T}^{ij}&=&\rho_{\rm B}v^i_{\rm B}v^j_{\rm B}=\mathcal{O}(\kappa^2\mathcal{R}^{-2}). \label{5:tijB}
\end{eqnarray}

\subsection{Ordering of the magnitude of the Newtonian perturbations}

We should note that there are two time-scales in the Newtonian system of our interest.
One is the timescale necessary to cross   
the length scale of the spacetime curvature radius with the speed of light,   
and another one is that to cross the system with the typical 
velocity $v_{\rm N}$ of the perturbations relative to the background; 
\begin{eqnarray}
 \mathcal{T}:=\frac{\mathcal{R}}{c}~~~~{\rm and}~~~~
 t_{\rm N}:=\frac{\ell_{\rm N}}{v_{\rm N}}.
\end{eqnarray}
By their definitions, we have
\begin{eqnarray}
 \frac{\mathcal{T}}{t_{\rm N}}=\frac{\epsilon}{\kappa}.
\end{eqnarray}
The typical dynamical timescale of the background is $\mathcal{T}$. 
By contrast, that of the perturbations 
will be equal to the shorter one of $\mathcal{T}$ and $t_{\rm N}$; if the effect of the 
spacetime curvature of the background, 
which appear as the anisotropic volume expansion of the background space, 
is more important than that of the self-gravity of the perturbations, 
the typical timescale of the evolution of the perturbations will be $\mathcal{T}$, whereas 
if it is not so important, the evolutions of the perturbations are determined by their 
self-gravity, and hence their typical dynamical timescale will be $t_{\rm N}$.  

We see from the above considerations that  
the order of magnitude of the time derivative of any quantity related to the perturbations  
is related to the order of magnitude of its spatial derivative in the manner,
\begin{equation}
\frac{\partial \psi}{\partial x^0}=\left\{
\begin{array}{ll}
\mathcal{O}\left(\epsilon~ \displaystyle{\frac{\partial \psi}{\partial x^i}}\right) &\quad {\rm for}~~\epsilon \gg \kappa, \\
&\\
\mathcal{O}\left(\kappa~ \displaystyle{\frac{\partial\psi }{\partial x^i}}\right)   &\quad {\rm for}~~\epsilon \ll   \kappa, \\
\end{array}\right.
\label{5:time}
\end{equation}
where $\psi$ is a representative quantity related to the perturbations. 
Hence, we will consider separately the three cases, 
$\epsilon \gg \kappa$, $\epsilon \simeq \kappa$ and $\epsilon \ll \kappa$, later.

It is not so difficult to find examples of those three cases in our universe.  
An example of the first case is the solar system.  The orbital speed of the earth 
is 30km/s, and the corresponding value of the parameter $\epsilon$ is 
about $10^{-4}$, whereas the mean spacetime curvature radius of the universe, or roughly speaking, 
the cosmological horizon scale is equal to 3Gpc, and corresponding value of the parameter $\kappa$ is 
about 1AU/3Gpc=2.0$\times10^{-15}$. An example of  the second case is the 
cluster of galaxies in which the velocity dispersion of galaxies is approximately 1000km/s 
and spatial scale is about 10Mpc, and hence 
we have $\epsilon \simeq \kappa \simeq 3.0\times 10^{-3}$.
An example of the third case is the structure on the BAO scale
whose velocity dispersion is approximately 
600km/s and the spatial scale is about $100 {\rm Mpc}$.
Thus we have $\epsilon \simeq 2\times 10^{-3}\ll \kappa \simeq 3\times 10^{-2}$.

We define the perturbation of 3-velocity field $v^i_{\rm N}$  
in the huge void universe model as 
\begin{equation}
v^i_{\rm N}:=v^i-\bar{v}^i,
 \label{5:dust1}
\end{equation}
where $v^i$ is the 3-velocity defined in the manner of Eq.~\eqref{3-veloity-def}, and 
$\bar{v}^i:=\bar{u}^i/\bar{u}^0$ is the background one given by
\begin{equation}
\bar{v}^i=v_{\rm B}^i+\mathcal{O}(\kappa^2).
\end{equation}  
As mentioned, we assume 
\begin{equation}
v_{\rm N}^i=\mathcal{O}(\epsilon). \label{5:vN1}
\end{equation} 
It is a little complicated to determine the order of magnitude of the density perturbation $\rho_{\rm N}$. 
The Newtonian dynamics holds up to the leading order 
in the sufficiently small domain of the background as can be seen from 
Eqs.~\eqref{5:cons1}--\eqref{5:poisson1}, and the Newtonian perturbations are also 
almost governed by the Newtonian dynamics by its definition. 
Hence the virial relation nearly holds in the whole 
system including both of the background and the Newtonian perturbations; 
\begin{eqnarray}
 \frac{G\rho\ell_{\rm N}^2}{c^2} \simeq \frac{\delta_{ij}v^i v^j}{c^2}.
 \label{5:virial}
\end{eqnarray}
Since the background quantities should satisfy the virial relation by only themselves, 
Eq.~\eqref{5:virial} leads to
\begin{equation}
 \frac{G\rho_{\rm N}\ell_{\rm N}^2}{c^2} \simeq  
 \delta_{ij}\left(2\frac{\bar{v}^i}{c}\frac{v^j_{\rm N}}{c}
 +\frac{v^i_{\rm N}}{c}\frac{v^j_{\rm N}}{c}\right).
\end{equation}
Hence, in the geometrized unit, we have
\begin{equation}
\rho_{\rm N}=\left\{
\begin{array}{ll}
\mathcal{O}(\epsilon^2\ell_{\rm N}^{-2})       &\quad {\rm for}~~\epsilon \gg \kappa, \\
\mathcal{O}(\kappa\epsilon\ell_{\rm N}^{-2})&\quad {\rm for}~~\epsilon \ll   \kappa. \\
\end{array}\right.,
\label{5:rhoN-order}
\end{equation}
We see from Eqs.~\eqref{rhoB-order} and \eqref{5:rhoN-order} that 
the order of magnitude of the density contrast is estimated at 
\begin{equation}
\frac{\rho_{\rm N}}{\bar{\rho}}=
\left\{
\begin{array}{ll}
\mathcal{O}(\kappa^{-2}\epsilon^2) \gg1      &\quad {\rm for}~~\epsilon \gg \kappa, \\
\mathcal{O}(\kappa^{-1}\epsilon) \ll1          &\quad {\rm for}~~\epsilon \ll   \kappa. \\
\end{array}\right.,
\label{5:delta1}
\end{equation}
Equation \eqref{5:delta1} implies that the absolute value of the density contrast is  much smaller 
than unity only if the inequality $\epsilon \ll \kappa$ holds. 
This is the same situation as that of the cosmological Newtonian approximation pointed 
out by Shibata and Asada~\cite{Shibata:1995dg}.

The order of magnitude of $h_{\mu\nu}^{\rm N}$ is determined through the Einstein equations, 
after fixing the gauge freedom of the perturbations. 
It should be noted that we have not yet fixed the coordinate system for 
the whole system composed of the background and perturbations, 
although the background is covered by the Fermi-normal coordinates. 
This means that we have not yet fixed 
the gauge freedom of the perturbations.

We fix the gauge freedom as follows. 
By the definition of the Newtonian perturbations, $|h_{\mu\nu}^{\rm N}|$ should be much smaller 
than unity, and we neglect the nonlinear terms with respect to $h^{\rm N}_{\mu\nu}$ 
since, as mentioned, we derive the basic equations for the Newtonian perturbations 
up to the leading order. 
The infinitesimal gauge transformation $x^\mu \rightarrow x^\mu +\xi^\mu$ causes a change in 
$h^{\rm N}_{\mu \nu}$ as 
\begin{equation}
 h^{\rm N}_{\mu \nu} 
 \to h^{\rm N}_{\mu \nu}+\partial_\mu \xi_\nu +\partial_\nu \xi_\mu 
 -2\bar{\Gamma}^\alpha_{\mu \nu}\xi_\alpha,
\end{equation}
where $\xi_\mu=\bar{g}_{\mu\nu}\xi^\nu$ is the smooth vector field, 
and $\bar{\Gamma}^\alpha_{\mu \nu}$ is the connection of the background metric $\bar{g}_{\mu\nu}$. 
Because of $\bar{\Gamma}^\alpha_{\mu \nu}=\mathcal{O}(\kappa^2)$, 
the leading order of the gauge transformation with respect to $\kappa$ is given by
\begin{eqnarray}
 h^{\rm N}_{\mu \nu}\to 
 h^{\rm N}_{\mu \nu}+\partial_\mu \xi_\nu +\partial_\nu \xi_\mu.
\end{eqnarray}
In this paper, in order to fix the gauge freedom,  we impose a condition 
which is the same form as the linearized harmonic condition in the Minkowski background;  
\begin{eqnarray}
 \partial^\mu\left(h^{\rm N}_{\mu \nu}
 -\frac{1}{2}\eta_{\mu\nu} \eta^{\alpha\beta}h_{\alpha\beta}^{\rm N}\right)=0,
 \label{5:harmonic}
\end{eqnarray}
where $\partial^\mu=\eta^{\mu\nu}\partial_\nu$.\footnote{It is worthwhile to notice that 
Eq.~\eqref{5:harmonic} 
coincides with the the harmonic condition $\partial_\mu(\sqrt{-g}g^{\mu\nu})=0$ 
up to the leading order with respect to $h^{\rm N}_{\mu\nu}$, 
only if the inequality $\epsilon\gg\kappa$ holds. 
Despite this fact, since, in the case of the Newtonian system, the slow motion approximation 
is valid by its definition, the condition \eqref{5:harmonic} uniquely fixes the gauge freedom 
by imposing a suitable boundary condition as in the case of the harmonic condition in the 
Minkowski background~\cite{wald}.} 
The Einstein tensor can be decomposed into the form, 
$G_{\mu \nu}=\bar{G}_{\mu \nu}+\delta G_{\mu \nu}$, 
where $\bar{G}_{\mu \nu}$ and $\delta G_{\mu \nu}$ denote 
the Einstein tensor of the background and that of the perturbations, respectively.
Up to the leading order, $\delta G_{\mu\nu}$ is given by  
\begin{eqnarray}
\delta G_{\mu \nu}=-\frac{1}{2}\eta^{\alpha\beta}\partial_\alpha\partial_\beta
\left(h^{\rm N}_{\mu \nu}
 -\frac{1}{2}\eta_{\mu\nu} \eta^{\rho\sigma}h_{\rho\sigma}^{\rm N}\right)
  \simeq -\frac{1}{2}\nabla^2\left(h^{\rm N}_{\mu \nu}
 -\frac{1}{2}\eta_{\mu\nu} \eta^{\rho\sigma}h_{\rho\sigma}^{\rm N}\right),
 \label{5:Gmn3}
\end{eqnarray}
where we neglected the terms differentiated with respect to the time coordinate 
in accordance with Eq.~\eqref{5:time}.

\subsection{The basic equations for the Newtonian perturbations}
In the previous subsection, we showed that 
the order of magnitude of the time derivatives of the Newtonian perturbations 
and that of the energy density of the Newtonian perturbations 
depend on which of $\epsilon$ or $\kappa$ is large. 
In this subsection, we derive the basic equations for the three cases, 
$\epsilon\gg\kappa$, $\epsilon\simeq\kappa$ and $\epsilon\ll\kappa$, separately. 

In order to derive the leading order of the basic equations for the Newtonian perturbations, 
we need only the conservation laws $\nabla_\mu T^{\mu\nu}=0$ and 
the $(00)$ component of the Einstein equations; 
\begin{eqnarray}
 \partial_0\rho +\partial_i(\rho v^i)&=&
 \rho\left[-\Gamma^\mu _{\mu 0}-\Gamma^\mu_{\mu i}v^i
 +\left(\Gamma^0_{00}+2\Gamma^0_{0i}v^i+\Gamma^0_{ij}v^iv^j\right)\right],
 \label{5:CON} \\
 \partial_0v^i+v^j\partial_jv^i&=&
 -\Gamma^i_{00}-2\Gamma^i_{0j}v^j-\Gamma^i_{jk}v^jv^k
 +\left(\Gamma^0_{00}+2\Gamma^0_{0j}v^j+\Gamma^0_{jk}v^jv^k\right)v^i,
 \label{5:EUL} \\
 G_{00}&=&
 8\pi T_{00},
 \label{5:POI}
\end{eqnarray}
The other  components of the Einstein equations give only  
the higher order corrections to the basic equations. 

\subsubsection{In the case of $\epsilon \gg\kappa $}
In the case of $\epsilon \gg\kappa$, the stress-energy tensor 
of the Newtonian perturbations up to the leading order are given by
\begin{eqnarray}
\delta T^{00}&=&\rho_{\rm N}=\mathcal{O}(\epsilon^2 \ell_{\rm N}^{-2}),  \label{5:SET1a}\\
\delta T^{0i} &=&\rho_{\rm N}v_{\rm N}^i=\mathcal{O}(\epsilon^3 \ell_{\rm N}^{-2}),  \label{5:SET1b}\\
\delta T^{ij}  &=&\rho_{\rm N}v_{\rm N}^i v_{\rm N}^j
=\mathcal{O}(\epsilon^4 \ell_{\rm N}^{-2}). \label{5:SET1c}
\end{eqnarray}
Then, by using Eqs.~\eqref{5:Gmn3} and \eqref{5:SET1a}--\eqref{5:SET1c}, 
the leading order of the Einstein equations for the Newtonian
perturbations, $\delta G_{\mu\nu}=8\pi \delta T_{\mu\nu}$, lead to 
\begin{eqnarray}
 \nabla^2\left(h^{\rm N}_{00}
 +\frac{1}{2}\eta^{\alpha\beta}h_{\alpha\beta}^{\rm N}\right)&=& -16\pi \rho_{\rm N}, 
 \label{5:E1a} \\
 \nabla^2\ h^{\rm N}_{0i}&=& -16\pi \rho_{\rm N}v_i^{\rm N}, 
 \label{5:E1b} \\
 \nabla^2\left(h^{\rm N}_{ij}
 -\frac{1}{2}\delta_{ij}\eta^{\alpha\beta}h_{\alpha\beta}^{\rm N}\right)&=&
 -16\pi \rho_{\rm N}v^{\rm N}_i v^{\rm N}_j,
 \label{5:E1c}
\end{eqnarray}
where we have introduced $v^{\rm N}_i:=v_{\rm N}^i$ for the notational consistency.
It is seen from Eqs.~\eqref{5:SET1a}--\eqref{5:SET1c} and \eqref{5:E1a}--\eqref{5:E1c}   
that the order of magnitude of each component of $h^{\rm N}_{\mu \nu}$ is 
\begin{eqnarray}
 h^{\rm N}_{00}=\mathcal{O}(\epsilon^2),~~~~
 h^{\rm N}_{0i}=\mathcal{O}(\epsilon^3),~~~~
 \delta^{ij}h^{\rm N}_{ij}=\mathcal{O}(\epsilon^2),~~~~
 h^{\rm N}_{ij}-\frac{1}{3}\delta_{ij}\delta^{kl}h^{\rm N}_{kl}=\mathcal{O}(\epsilon^4).
 \label{5:hN1}
\end{eqnarray}

By subtracting the background equations  \eqref{5:cons1}, 
\eqref{5:euler1} and \eqref{5:poisson1} from Eqs.~\eqref{5:CON}--\eqref{5:POI}, 
we obtain, in accordance with the ordering~\eqref{5:time}, 
\eqref{5:vN1}, \eqref{5:rhoN-order} and \eqref{5:hN1}, 
the leading order of the basic equations governing the Newtonian perturbations as
\begin{eqnarray}
 \partial_0\rho_{\rm N}+\partial_i\left(\rho_{\rm N}v_{\rm N}^i\right)&=& 0,
 \label{5:f1a} \\
 \partial_0 v_{\rm N}^i+v_{\rm N}^j \partial_j v_{\rm N}^i
 &=& -\partial^i \Phi_{\rm N},
 \label{5:f1b} \\
 \nabla^2 \Phi_{\rm N}
 &=& 4\pi \rho_{\rm N}.
 \label{5:f1c}
\end{eqnarray}
where 
\begin{eqnarray}
 \Phi_{\rm N}:=-h^{\rm N}_{00}/2
\end{eqnarray}
corresponds to the Newtonian 
gravitational potential, and $\partial^i:=\partial_i$. Equations~\eqref{5:f1a}--\eqref{5:f1c}
are the same as the basic equations for the gravitating dust in the framework of 
the Newtonian theory of gravity. 
The effects of the tidal field of the background spacetime do not appear up to the leading order. 

\subsubsection{In the case of $\epsilon \simeq\kappa $}

In this case, the number of parameters characterizing the system becomes only one; we   
replace $\kappa$ by $\epsilon$. The leading order of the stress-energy tensor is given by
\begin{eqnarray}
\delta T^{00}&=&\rho_{\rm N}=\mathcal{O}(\epsilon^2 \ell_{\rm N}^{-2}),  \label{5:SET2a}\\
\delta T^{0i} &=&\rho_{\rm B}v_{\rm N}^i+\rho_{\rm N}\left(v_{\rm B}^i+v_{\rm N}^i\right)
=\mathcal{O}(\epsilon^3 \ell_{\rm N}^{-2}),  \label{5:SET2b}\\
\delta T^{ij}  &=&\rho_{\rm B}\left(2v_{\rm B}^{(i}v_{\rm N}^{j)}+v_{\rm N}^i v_{\rm N}^j\right)
+\rho_{\rm N}\left(v_{\rm B}^i v_{\rm B}^j+2v_{\rm B}^{(i} v_{\rm N}^{j)}+v_{\rm N}^i v_{\rm N}^j\right)
=\mathcal{O}(\epsilon^4 \ell_{\rm N}^{-2}). \label{5:SET2c}
\end{eqnarray}
Since the order of magnitude of the stress-energy tensor is the same as Eqs.~\eqref{5:SET1a}--\eqref{5:SET1c}, 
the orders of magnitudes of the metric perturbations are given by Eq.~\eqref{5:hN1}. 

Then the basic equations for the Newtonian perturbations up to the leading order are derived 
from Eqs.~\eqref{5:CON}--\eqref{5:POI}, and we have
\begin{eqnarray}
\partial_0\rho_{\rm N}+\partial_i\left(\rho_{\rm N}v_{\rm N}^i\right)
+\partial_i\left(\rho_{\rm N}v_{\rm B}^i\right)
+\partial_i\left(\rho_{\rm B}v_{\rm N}^i\right)&=&0, \label{5:f2a}\\
 \partial_0 v_{\rm N}^i+v_{\rm N}^j \partial_j v_{\rm N}^i
+v_{\rm N}^j \partial_j v_{\rm B}^i+v_{\rm B}^j \partial_j v_{\rm N}^i 
 &=& -\partial^i \Phi_{\rm N},
 \label{5:f2b} \\
 \nabla^2 \Phi_{\rm N}
 &=& 4\pi \rho_{\rm N}.
 \label{5:f2c}
\end{eqnarray}
Equations~\eqref{5:f2a}--\eqref{5:f2c} imply 
that the effects of the background spacetimes appear in the third and forth terms 
in the left hand sides of Eqs.~\eqref{5:f2a} and \eqref{5:f2b} through $\rho_{\rm B}$ and $v_{\rm B}^i$. 
Here it is worthwhile to notice that the background velocity field $v_{\rm B}^i$ 
is generated by the background gravitational field $\Phi_{\rm B}$, or equivalently 
the background tidal field $h_{00}^{\rm B}$ by Eq.~(\ref{5:local1-phi}), through Eq.~(\ref{5:euler1}).
It is expected that the evolutions of Newtonian perturbations reflect the tidal field of the background 
huge void in the case of $\epsilon\simeq\kappa$.

\subsubsection{In the case of $\epsilon \ll \kappa $}

In this case, the ordering for the Newtonian perturbations is quite different from 
that in the previous two cases. The stress-energy tensor up to the leading order is given by 
\begin{eqnarray}
\delta T^{00}&=&\rho_{\rm N}=\mathcal{O}(\epsilon\kappa \ell_{\rm N}^{-2}),  \label{5:SET3a}\\
\delta T^{0i} &=&\rho_{\rm B}v_{\rm N}^i+\rho_{\rm N}v_{\rm B}^i
=\mathcal{O}(\epsilon\kappa^2 \ell_{\rm N}^{-2}),  \label{5:SET3b}\\
\delta T^{ij}  &=& 2\rho_{\rm B}v_{\rm B}^{(i}v_{\rm N}^{j)}
+\rho_{\rm N}v_{\rm B}^i v_{\rm B}^j
=\mathcal{O}(\epsilon\kappa^3 \ell_{\rm N}^{-2}). \label{5:SET3c}
\end{eqnarray}
By using Eqs.~\eqref{5:Gmn3} and \eqref{5:SET3a}--\eqref{5:SET3c},
we obtain the leading order of the 
equations for the metric perturbations as  
\begin{eqnarray}
 \nabla^2\bar{h}^{\rm N}_{00}&=& -16\pi \rho_{\rm N}, 
 \label{5:E2a} \\
 \nabla^2\bar{h}^{\rm N}_{0i}&=& -16\pi \left(\rho_{\rm B}v_{\rm N}^i+\rho_{\rm N}v_{\rm B}^i\right), 
 \label{5:E2b} \\
 \nabla^2\bar{h}^{\rm N}_{ij}&=& -16\pi \left(2\rho_{\rm B} v_{\rm B}^{(i}v_{\rm N}^{j)}
 +\rho_{\rm N}v_{\rm B}^i v_{\rm B}^j\right).
 \label{5:E2c}
\end{eqnarray}
By using Eqs.~\eqref{5:SET3a}--\eqref{5:SET3c} and \eqref{5:E2a}--\eqref{5:E2c}, 
the order of magnitude of each component of $h_{\mu\nu}^{\rm N}$ is given by 
\begin{eqnarray}
 h^{\rm N}_{00}=\mathcal{O}(\epsilon\kappa),~~~~
 h^{\rm N}_{0i}=\mathcal{O}(\epsilon\kappa^2),~~~~
 \delta^{ij}h^{\rm N}_{ij}=\mathcal{O}(\epsilon\kappa),~~~~
 h^{\rm N}_{ij}-\frac{1}{3}\delta_{ij}\delta^{kl}h^{\rm N}_{kl}=\mathcal{O}(\epsilon\kappa^3).
\label{5:hN2}
\end{eqnarray}
We should note that the ordering of $h_{\mu \nu}$ in Eq.~\eqref{5:hN2} is quite different from that in Eq.~\eqref{5:hN1}. 

The basic equations for the Newtonian perturbations up to the leading order are derived 
from Eqs.~\eqref{5:CON}--\eqref{5:POI}, and we have
\begin{eqnarray}
\partial_0\rho_{\rm N}+\partial_i\left(\rho_{\rm N}v_{\rm B}^i\right)
+\partial_i\left(\rho_{\rm B}v_{\rm N}^i\right)&=&0, \label{5:f3a}\\
 \partial_0 v_{\rm N}^i
+v_{\rm N}^j \partial_j v_{\rm B}^i+v_{\rm B}^j \partial_j v_{\rm N}^i 
 &=& -\partial^i \Phi_{\rm N},
 \label{5:f3b} \\
 \nabla^2 \Phi_{\rm N}
 &=& 4\pi \rho_{\rm N}.
 \label{5:f3c}
\end{eqnarray}

\section{Analysis of linear Newtonian perturbations}\label{sec4}

In this section, we study the dynamical behavior of Newtonian perturbations whose 
amplitude is so small that the linear approximation is available. As 
shown in the previous section, the parameter 
$\epsilon$ is necessarily much smaller than the parameter $\kappa$, and hence 
the basic equations are given by Eqs.~\eqref{5:f3a}--\eqref{5:f3c} which have 
already been linearized.

It is worthwhile to notice that, in general, it is very difficult to analytically study the anisotropic linear perturbations in the background LTB solution. Hence, it is a very non-trivial subject to study analytically 
the Newtonian perturbations in the huge void universe model based on the LTB solution, even if 
their amplitude is so small that the linear approximation is applicable.

\subsection{Basic equations for the linear Newtonian perturbations}

Hereafter we denote the density contrast by $\delta_{\rm N}$;
\begin{equation}
\delta_{\rm N}:=\frac{\rho_{\rm N}}{\bar{\rho}}=\frac{\rho_{\rm N}}{\rho_{\rm B}}
\left[1+\mathcal{O}(\kappa)\right].
\label{delta-def}
\end{equation}
By using Eqs.~\eqref{5:local1-v} and \eqref{delta-def}, Eqs.~\eqref{5:f3a}--\eqref{5:f3c} are 
rewritten in the following forms;
\begin{eqnarray}
 \left(\partial_0+v_{\rm B}^j\partial_j\right)
 \delta_{\rm N}
 +\partial_jv^j_{\rm N}
 &=& 0,
 \label{5:lin1} \\
 &&\cr
 \left(\partial_0+v_{\rm B}^j\partial_j\right)v^i_{\rm N}
 +H^i{}_jv^j_{\rm N}
 &=& -\partial^i\Phi_{\rm N},
 \label{5:lin2} \\
 &&\cr
 \nabla^2 \Phi_{\rm N}&=&
 4\pi \rho_{\rm B}\delta_{\rm N}.
 \label{5:lin3}
\end{eqnarray}

We introduce the following kinematical variables of the Newtonian perturbations;  
\begin{eqnarray}
\Theta_{\rm N}&:=&\delta^{ij}\partial_j v^{\rm N}_i, \\
\sigma^{\rm N}_{ij}&:=&\partial_{(j}v^{\rm N}_{i)}-\frac{1}{3}\delta_{ij}\Theta_{\rm N}, \\
\omega^{\rm N}_{ij}&:=&\partial_{[j}v^{\rm N}_{i]}.
\end{eqnarray}
We call $\Theta_{\rm N}$, $\sigma^{\rm N}_{ij}$ and $\omega^{\rm N}_{ij}$  
the expansion, shear and vorticity of the Newtonian perturbations,  
respectively. We also introduce the similar variables to the above ones but
related to the background 3-velocity field;
\begin{eqnarray}
\Theta_{\rm B}&:=&\delta^{ij}\partial_j v^{\rm B}_i, \label{exp-B}\\
\sigma^{\rm B}_{ij}&:=&\partial_{(j}v^{\rm B}_{i)}-\frac{1}{3}\delta_{ij}\Theta_{\rm B}, \label{shear-B}\\
\omega^{\rm B}_{ij}&:=&\partial_{[j}v^{\rm B}_{i]}. \label{rot-B}
\end{eqnarray}
We call $\Theta_{\rm B}$, $\sigma^{\rm B}_{ij}$ and $\omega^{\rm B}_{ij}$  
the expansion, shear and vorticity of the background, respectively.
By using Eqs.~\eqref{5:matrix1} and \eqref{5:local1-v},
the kinematical variables of the background \eqref{exp-B}--\eqref{rot-B} are given by
\begin{eqnarray}
 \Theta_{\rm B}(x^0)=H_\parallel^{\rm B}(x^0)+2H_\bot^{\rm B}(x^0),
 \label{5:expB}
\end{eqnarray}
\begin{eqnarray}
 \sigma^{\rm B}_{ij}(x^0)=
 \left(
 \begin{array}{ccc}
 2\sigma_B & 0 & 0 \\
 0 & -\sigma_{\rm B} & 0 \\
 0 & 0 & -\sigma_{\rm B}
\end{array} 
 \right),
 \label{5:shearB}
\end{eqnarray}
and
\begin{eqnarray}
 \omega^{\rm B}_{ij}(x^0)=0,\label{5:vortB}
\end{eqnarray}
where
\begin{equation}
\sigma_{\rm B}(x^0):=\frac{1}{3}\left[H_\parallel^{\rm B}(x^0)-H_\bot^{\rm B}(x^0)\right] .
\end{equation}

We introduce the Lagrangian coordinates $q^\mu$ with respect to the background, which are defined as
\begin{equation}
x^0=q^0~~~~~{\rm and}~~~~~~x^i=q^i+X^i(q^0),
\end{equation}
where $X^i(q^0)$ is defined by the following differential equation;
\begin{equation}
\frac{\partial X^i}{\partial q^0}(q^0)=v_B^i(x^0).
\end{equation}
We have
\begin{equation}
\frac{\partial}{\partial q^0}=\partial_0+v_{\rm B}^j\partial_j~~~~~{\rm and}~~~~~
\frac{\partial}{\partial q^i}=\partial_i. 
\end{equation}
Then, by using these background Lagrangian coordinates, 
Eqs.~\eqref{5:lin1}--\eqref{5:lin3} lead to the following equations for the density contrast 
$\delta_{\rm N}$, the kinematical
variables, $\Theta_{\rm N}$, $\sigma^{\rm N}_{ij}$ and $\omega^{\rm N}_{ij}$ and 
the gravitational potential $\Phi_{\rm N}$, as
\begin{eqnarray}
 \frac{\partial \delta_{\rm N}}{\partial q^0}&=& -\Theta_{\rm N}
 \label{5:L1a} \\
 \frac{\partial\Theta_{\rm N}}{\partial q^0}&=& 
 -\frac{2}{3}\Theta_{\rm B}\Theta_{\rm N}-2\delta^{ij}\delta^{kl}\sigma^{\rm B}_{ik}\sigma^{\rm N}_{jl}
 -4\pi \rho_{\rm B}\delta_{\rm N},
 \label{5:L1b} \\
 \frac{\partial\sigma^{\rm N}_{ij}}{\partial q^0}&=&
 -\frac{2}{3}\Theta_{\rm B}\sigma^{\rm N}_{ij}
 -\frac{2}{3}\Theta_{\rm N}\sigma^{\rm B}_{ij}
 -2\left(\delta^{kl}\sigma^{\rm B}_{k(i}\sigma^{\rm N}_{j)l}
 -\frac{1}{3}\delta_{ij}\delta^{kl}\delta^{mn}\sigma^{\rm B}_{km}\sigma^{\rm N}_{ln}\right) \cr
  &-&\left(\frac{\partial^2}{\partial q^i\partial q^j}-\frac{1}{3}\delta_{ij}\nabla^2_q\right)\Phi_{\rm N},
  \label{5:L1c}\\
 \frac{\partial \omega^{\rm N}_{ij}}{\partial q^0}&=&
 -\frac{2}{3}\Theta_{\rm B}\omega^{\rm N}_{ij}
 +2\delta^{kl}\sigma^{\rm B}_{k[i}\omega^{\rm N}_{j]l},
 \label{5:L1d} \\
 \nabla_q^2 \Phi_{\rm N}&=& 4\pi \rho_{\rm B}\delta_{\rm N},
 \label{5:L1e}
\end{eqnarray}
where $\nabla^2_q$ is the flat Laplacian operator with respect to $q^i$.

Equations~\eqref{5:L1a}--\eqref{5:L1e} are linear partial differential equations with respect to 
$q^{\mu}$.  The remarkable feature of these equations is that all of the coefficients of the 
perturbation variables depend on only $q^0$. Hence, the Fourier transformation 
is very useful to solve these equations.
The Fourier transforms of the Newtonian perturbations 
are denoted by the symbols with a tilde. For example, the density contrast $\delta_{\rm N}$ 
is written in the form
\begin{eqnarray}
 \delta_{\rm N}(q^0,q^i)=\int \frac{d^3k}{(2\pi)^{3/2}}e^{i k_jq^j}\tilde{\delta}_{\rm N}(q^0,k^i).
 \label{5:Fourier}
\end{eqnarray}
From Eqs.~\eqref{5:L1a}--\eqref{5:L1e}, we obtain the equations for the Fourier transforms as
\begin{eqnarray}
 \frac{\partial \tilde{\delta}_{\rm N}}{\partial q^0}&=& -\tilde{\Theta}_{\rm N}
 \label{5:L2a} \\
 \frac{\partial\tilde{\Theta}_{\rm N}}{\partial q^0}&=& 
 -\frac{2}{3}\Theta_{\rm B}\tilde{\Theta}_{\rm N}
 -2\delta^{ij}\delta^{kl}\sigma^{\rm B}_{ik}\tilde{\sigma}^{\rm N}_{jl}
 -4\pi \rho_{\rm B}\tilde{\delta}_{\rm N},
 \label{5:L2b} \\
 \frac{\partial\tilde{\sigma}^{\rm N}_{ij}}{\partial q^0}&=&
 -\frac{2}{3}\Theta_{\rm B}\tilde{\sigma}^{\rm N}_{ij}
 -\frac{2}{3}\tilde{\Theta}_{\rm N}\sigma^{\rm B}_{ij}
 -2\left(\delta^{kl}\sigma^{\rm B}_{k(i}\tilde{\sigma}^{\rm N}_{j)l}
 -\frac{1}{3}\delta_{ij}\delta^{kl}\delta^{mn}\sigma^{\rm B}_{km}\tilde{\sigma}^{\rm N}_{ln}\right) \cr
 &+&\left(k_i k_j-\frac{1}{3}k^2\delta_{ij}\right)\tilde{\Phi}_{\rm N},
 \label{5:L2c}\\
 \frac{\partial \tilde{\omega}^{\rm N}_{ij}}{\partial q^0}&=&
 -\frac{2}{3}\Theta_{\rm B}\tilde{\omega}^{\rm N}_{ij}
 +2\delta^{kl}\sigma^{\rm B}_{k[i}\tilde{\omega}^{\rm N}_{j]l},
 \label{5:L2d} \\
 -k^2\tilde{ \Phi}_{\rm N}&=& 4\pi \rho_{\rm B}\tilde{\delta}_{\rm N},
 \label{5:L2e}
\end{eqnarray}
where $k^2=\delta_{ij}k^ik^j$. 
Equations~\eqref{5:L2a}--\eqref{5:L2e} form a set of the ordinary 
differential equations with respect to $q^0$. It should be noted that 
each Fourier mode is decoupled with the other modes, and this result 
is very different from the case of relativistic linear perturbations 
in the LTB solution(see, for example~\cite{Gerlach:1980}).

We have not yet fixed the spatial boundary condition which is necessary 
for obtaining an unique solution of the Poisson equation  \eqref{5:L1e}. 
However it should be noted that once we know the time dependence of 
$\tilde{\delta}_{\rm N}$, 
$\tilde{\sigma}^{\rm N}_{ij}$, $\tilde{\omega}^{\rm N}_{ij}$ and 
$\tilde{\Phi}_{\rm N}$ with all $k^i$,  
we can construct solutions with any boundary conditions by superposing them  
with appropriate weights, by virtue of the completeness of $e^{ik_jq^j}$.

\subsection{Evolution of vorticity}
First of all, we consider the vorticity of the Newtonian perturbations $\omega_{ij}^{\rm N}$. 
The equation for the vorticity \eqref{5:L2d} is decoupled from other equations. 
This is because the vorticity of the background $\omega_{ij}^{\rm B}$ vanishes 
as shown in Eq.~\eqref{5:vortB}.
By using Eqs.~\eqref{5:expB}--\eqref{5:shearB},
Eq.~\eqref{5:L2d} is rewritten in the form
\begin{eqnarray}
 \frac{\partial\tilde{\omega}_{12}^{\rm N}}{\partial q^0}
 &=&-\left[H_\parallel^{\rm B}(q^0)+H_\bot^{\rm B}(q^0)\right]\tilde{\omega}_{12}^{\rm N},
 \label{5:vort12} \\
 \frac{\partial\tilde{\omega}_{13}^{\rm N}}{\partial q^0}
 &=&-\left[H_\parallel^{\rm B}(q^0)+H_\bot^{\rm B}(q^0)\right]\tilde{\omega}_{13}^{\rm N},
 \label{5:vort13} \\
 \frac{\partial\tilde{\omega}_{23}^{\rm N}}{\partial q^0}
 &=&-2H_\bot^{\rm B}(q^0)\tilde{\omega}_{23}^{\rm N}.
 \label{5:vort23}
\end{eqnarray}
By solving Eqs.~\eqref{5:vort12}--\eqref{5:vort23}, we obtain
\begin{eqnarray}
 \tilde{\omega}^{\rm N}_{12}(q^0,k^i)&=&\frac{C_{12}(k^i)}{a_\parallel^{\rm B}(q^0)a_\bot^{\rm B}(q^0)}, 
  \label{5:vort1} \\
 \tilde{\omega}^{\rm N}_{13}(q^0,k^i)&=&\frac{C_{13}(k^i)}{a_\parallel^{\rm B}(q^0)a_\bot^{\rm B}(q^0)}, \\
\tilde{\omega}^{\rm N}_{23}(q^0,k^i)&=&\frac{C_{23}(k^i)}{\left[a_\bot^{\rm B}(q^0)\right]^2},
 \label{5:vort3}
\end{eqnarray}
where $C_{12}$, $C_{13}$ and $C_{23}$ are arbitrary functions of $k^i$, and 
the scale factors, $a_\parallel^{\rm B}$ and $a_\bot^{\rm B}$, are defined by the differential equations as,
\begin{equation}
\frac{d\ln a_\parallel^{\rm B}(q^0)}{dq^0}=H_\parallel^{\rm B}(q^0)~~~~{\rm and}~~~~
\frac{d\ln a_\bot^{\rm B}(q^0)}{dq^0}=H_\bot^{\rm B}(q^0).
\end{equation}

From the solutions~\eqref{5:vort1}--\eqref{5:vort3}, 
we can see that the vorticity of the Newtonian perturbations decays as time goes on, 
since both scale factors, $a_\parallel^{\rm B}$ and $a_\bot^{\rm B}$, grow as time goes on,  
in the case of the huge void universe model.
It is well known that the linear vorticity decays as 
$\omega_{ij}\propto a^{-2}$ in the case of homogeneous and isotropic universe model, 
where $a$ is the scale factor of this model.
Since the time dependence of scale factors of the huge void universe model 
may be significantly different from that of the homogeneous and isotropic universe model,
the evolution of the linear vorticity in the huge void universe model 
may significantly differ from that in the homogeneous and isotropic universe model.

\subsection{Evolution of density contrast}
In contrast to the equation for the vorticity of the Newtonian perturbations, 
the equations for the other variables, 
$\delta_{\rm N}$, $\Theta_{\rm N}$, $\sigma_{ij}^{\rm N}$ and $\Phi_{\rm N}$, that is,  
Eqs.~\eqref{5:L2a}--\eqref{5:L2c} and \eqref{5:L2e}, are coupled to each other.
We solve these coupled equations numerically and study the growth of density contrast $\delta_{\rm N}$.

In order to see the evolution of $\tilde{\delta}_{\rm N}$, we need  
Eqs.~\eqref{5:L2a}, \eqref{5:L2b} and the 1-1 components of Eqs.~\eqref{5:L2c} and \eqref{5:L2e},
but the other equations are not necessary. 
By using Eq.~\eqref{5:shearB}, the system of the equations 
necessary for studying the density contrast is obtained in the following form; 
\begin{eqnarray}
 \frac{\partial\tilde{\delta}_{\rm N}}{\partial q^0}&=&-\tilde{\Theta}_{\rm N},
 \label{5:sca1} \\
 \frac{\partial\tilde{\Theta}_{\rm N}}{\partial q^0}&=&-\frac{2}{3}\Theta_{\rm B}\tilde{\Theta}_{\rm N}
 -6\sigma_{\rm B}\tilde{\sigma}^{\rm N}_{11}
 -4\pi \rho_{\rm B}\tilde{\delta}_{\rm N}, 
 \label{5:sca2} \\
 \frac{\partial\tilde{\sigma}^{\rm N}_{11}}{\partial q^0}&=&-\frac{4}{3}\sigma_{\rm B}\tilde{\Theta}_{\rm N}
 -\frac{2}{3}\Theta_{\rm B}\tilde{\sigma}^{\rm N}_{11}
 -2\sigma_{\rm B}\tilde{\sigma}^{\rm N}_{11}
 -4\pi \rho_{\rm B}\left(\mu^2-\frac{1}{3}\right)\tilde{\delta}_{\rm N},
 \label{5:sca3}
\end{eqnarray}
where $\mu:=k_1/k$.
Once $\delta_{\rm N}$, $\Theta_{\rm N}$ and $\sigma^{\rm N}_{11}$ are obtained 
by solving Eqs.~\eqref{5:sca1}--\eqref{5:sca3},
other components of the shear $\sigma^{\rm N}_{ij}$ and the gravitational potential $\Phi_{\rm N}$
can be determined by solving Eqs.~\eqref{5:L2c} and \eqref{5:L2e}.

We assume that the huge void universe model has the uniform Big-Bang time 
and approaches the Einstein de-Sitter universe model for $r\to\infty$. This assumption is 
consistent to the inflationary scenario. By this assumption, $\sigma_{\rm B}$ almost vanishes, 
and $\Theta_{\rm B}$ and $\rho_B$ behave as those in the Einstein-de Sitter universe model,  
much before the huge void structure becomes prominent. 
Hence, Eqs.~\eqref{5:sca1} and \eqref{5:sca2} in the early stage lead to the well known equation for the 
density contrast in the Einstein-de Sitter universe model,
\begin{equation}
\frac{\partial^2 \tilde{\delta}_{\rm N}}{\partial(q^0)^2}
+2H\frac{\partial \tilde{\delta}_{\rm N}}{\partial q^0}-4\pi\rho_{\rm EdS}\tilde{\delta}_{\rm N}=0,
\label{delta-EdS}
\end{equation}
where 
\begin{equation}
H=\frac{2}{3q^0}~~~~{\rm and}~~~~\rho_{\rm EdS}\propto \frac{1}{(q^0)^{2}}.
\end{equation}
There are two independent solutions 
of Eq.~\eqref{delta-EdS}; $D_+=(q^0)^{2/3}$ and $D_-=(q^0)^{-1}$. The solutions proportional to $D_+$ are  
called the growing modes, whereas those proportional to $D_-$ are called the decaying modes. 
Since the decaying modes are observationally unimportant, hereafter, we neglect them. 
Then, the solutions of Eqs.~\eqref{5:sca1}--\eqref{5:sca3} in the early stage are given by 
\begin{eqnarray}
\tilde{\delta}_{\rm N}&\simeq&\delta(k^i)\left(\frac{q^0}{q^0_{\rm i}}\right)^{2/3}, \\
\tilde{\Theta}_{\rm N}&=&-\frac{\partial \tilde{\delta}_{\rm N}}{\partial q^0}\simeq
-\frac{2}{3q^0_{\rm i}}\delta(k^i)\left(\frac{q^0}{q^0_{\rm i}}\right)^{-1/3}, \\
\tilde{\sigma}^{\rm N}_{11}&\simeq&-\left(\mu^2-\frac{1}{3}\right)\frac{\partial\tilde{\delta}_{\rm N}}{\partial q^0}
\simeq-\frac{2}{3q^0_{\rm i}}\left(\mu^2-\frac{1}{3}\right)\delta(k^i)\left(\frac{q^0}{q^0_{\rm i}}\right)^{-1/3},
\end{eqnarray}
where $q^0=q^0_{\rm i}$ denotes the Lagrangian time at which we set initial data for 
Eqs.~\eqref{5:sca1}--\eqref{5:sca3}. 
Hence we fix the initial conditions for Eqs.~\eqref{5:sca1}--\eqref{5:sca3} as follows; 
\begin{eqnarray}
 \tilde{\delta}_{\rm N}(q^0_{\rm i},k^i)&=&\delta(k^i),
 \label{5:initial1} \\
 \tilde{\theta}_{\rm N}(q^0_{\rm i},k^i)&=&-\frac{2}{3q^0_{\rm i}}\delta(k^i),
 \label{5:initial2} \\
 \tilde{\sigma}^{\rm N}_{11}(q^0_{\rm i},k^i)&=&-\frac{2}{3q^0_{\rm i}}
 \left(\mu^2-\frac{1}{3}\right)\delta(k^i),
 \label{5:initial3}
\end{eqnarray}
where $\delta(k^i)$ is an arbitrary function of $k^i$. 
We assume $\delta(k^i)$ to have a stochastic property
\begin{eqnarray}
 \langle \delta(k^i)\delta({k'}^i)\rangle
 =P(k)\delta_{\rm D}^3(k^i-{k'}^i),
\end{eqnarray}
where $\delta_{\rm D}$ denotes Dirac's delta function,
and $P(k)$ represents the power spectrum. 
From Eqs.~\eqref{5:sca1}--\eqref{5:sca3} and the initial conditions~\eqref{5:initial1}--\eqref{5:initial3},
the density contrast is written in the form
\begin{eqnarray}
 \tilde{\delta}_{\rm N}(q^0,k^i)~=~D(q^0,\mu)\delta(k^i).
 \label{5:growth1}
\end{eqnarray}
where $D(q^0,\mu)$ behaves as $D\simeq (q^0/q^0_{\rm i})^{2/3}$ in the early stage and hereafter 
is called the growth factor of the density contrast. 
Since the evolution of the shear $\sigma^{\rm N}_{11}$ depends on $\mu$, 
and the density contrast $\delta_{\rm N}$ couples to the shear $\sigma^{\rm N}_{11}$ 
through the non-vanishing $\sigma_{\rm B}$, 
the dependence of the growth factor $D$ on $\mu$ comes  
from the anisotropy of the volume expansion.

Here, we should recall that 
the domain we consider is covered by the Fermi-normal coordinates whose fundamental timelike geodesic 
agrees with the curve specified by $r'=r'_{\rm o}$ with $\theta'$ and $\phi'$ fixed. 
This fact implies that the background quantities, $\rho_{\rm B}$, 
$H_\parallel^{\rm B}$ and $H_\bot^{\rm B}$ depend on the parameter $r'_{\rm o}$;   
$\rho_{\rm B}(q^0)=\bar{\rho}(t',r'_{\rm o})$, 
$H^{\rm B}_{\parallel}(q^0)=H_{\parallel}(t', r'_{\rm o})$ and 
$H^{\rm B}_{\bot}(q^0)=H_{\bot}(t';r'_{\rm o})$.
Thus, the arguments of the density contrast $\tilde{\delta}_{\rm N}$ and the 
growth factor $D$ should be revised as follows;  
\begin{eqnarray}
 \tilde{\delta}_{\rm N}(q^0,k^i;r_{\rm o}')~=~D(q^0,\mu;r_{\rm o}')\delta(k^i).
 \label{5:growth2}
\end{eqnarray}
The dependence of the growth factor on $r_{\rm o}$ comes from the radial 
inhomogeneities of the background huge void universe model.

In order to study the evolution of growth factor $D(q^0,\mu;r_{\rm o}')$,
we consider a toy version of the huge void universe model which is,  
hereafter, called the {\it toy background model}.
In order to fix the toy background model, 
we introduce the density-parameter function $\Omega(r')$ defined as 
\begin{equation}
\Omega(r'):=\frac{8\pi}{H_\bot^2(t_0',r')R^3(t_0',r')}\int_0^{r'}dx\bar{\rho}(t_0',x)
R^2(t_0',x)\partial_xR(t_0',x),
\end{equation}
where $t'=t'_0$ is the present time. 
We assume  
\begin{eqnarray}
 \Omega(r')=\Omega_{\rm out}-\left(\Omega_{\rm out}-\Omega_{\rm in}\right)e^{-\frac{r'^2}{2\sigma^2}},
 \label{5:voidmodel}
\end{eqnarray}
where $\Omega_{\rm out}=1.0$, $\Omega_{\rm in}=0.3$ and $\sigma=0.5t'_0$.

\begin{figure}[htbp]
 \begin{center}
 \includegraphics[width=8cm,clip]{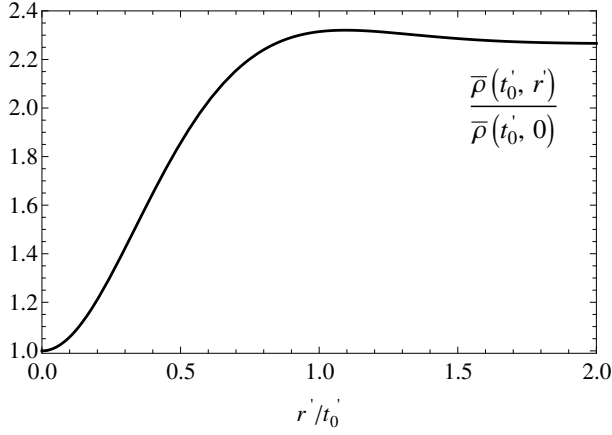}
 \end{center}
 \caption{
 The energy density $\bar{\rho}$ on the spacelike hypersurface specified by $t'=t_0'$ 
 in the toy background model is depicted 
 as a function of the radial coordinate $r'$.
 }
 \label{5:fig_TOY1}
\end{figure}

In Fig.~\ref{5:fig_TOY1}, we depict the energy density normalized by its value at $r'=0$ 
on the spacelike hypersurface specified by the present time $t'=t_0'$ of this toy background model 
as a function of the radial coordinate. 
It is seen from this figure that the toy background model has a non-linear void structure whose size is about $0.7t_0'$ at $t'=t'_0$.
The vicinity of the symmetry center $r'=0$ is well approximated by the dust filled FLRW universe model 
with the density parameter $\Omega_{\rm M}=0.3$, whereas the asymptotic region 
agrees with the Einstein-de Sitter universe model.

\begin{figure}[htbp]
 \begin{center}
 \includegraphics[width=8cm,clip]{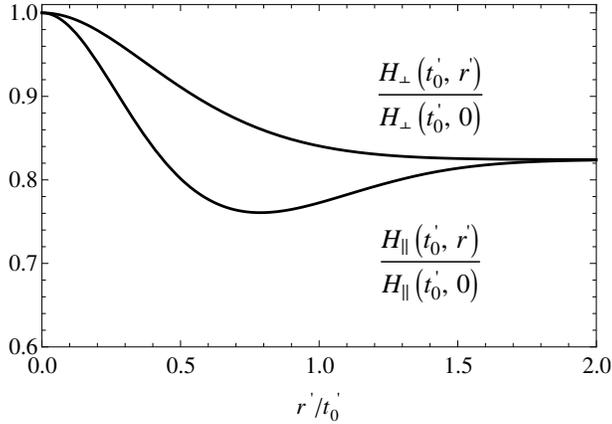}
 \end{center}
 \caption{
 The Hubble functions, $H_\parallel(t',r')$ and $H_\bot(t',r')$, on the spacelike hypersurface specified by 
 $t'=t_0'$ in the toy background model as functions of the radial coordinate.
 }
 \label{5:fig_TOY2}
\end{figure}

In Fig.~\ref{5:fig_TOY2}, the Hubble functions, $H_\parallel(t',r')$ and $H_\bot(t',r')$, 
normalized by their values at $r'=0$ 
on the spacelike hypersurface specified by $t'=t'_0$ 
in the toy background model are depicted as functions of the radial coordinate.
It is seen from this figure that the Hubble functions take their maximal values 
at the center $r'=0$, since the energy density $\bar{\rho}$ takes its minimal value at $r'=0$.

\begin{figure}[htbp]
 \begin{center}
 \includegraphics[width=8cm,clip]{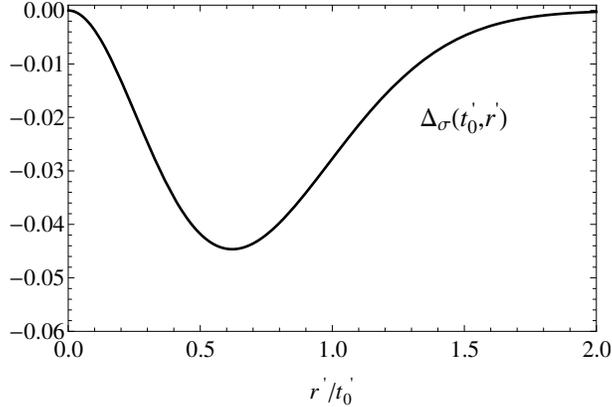}
 \end{center}
 \caption{
The normalized shear, $\Delta_{\rm \sigma}(t',r')$, of the toy background model as a function of $r'$ 
on the spacelike hypersurface specified by $t'=t_0'$.
 }
 \label{5:fig_TOY3}
\end{figure}

In Fig.~\ref{5:fig_TOY3}, we plot the normalized shear, $\Delta_{\rm \sigma}(t',r')$, defined as
\begin{eqnarray}
 \Delta_{\rm \sigma}(t',r'):=\frac{H_{\parallel}(t',r')-H_\bot(t',r')}{H_{\parallel}(t',r')+2H_\bot(t',r')},
\end{eqnarray}
on the spacelike hypersurfaces specified by $t'=t_0'$ in the toy background model as a function of $r'$.
We see from this figure that the minimal value of $\Delta_{\rm \sigma}(t_0',r')$ 
appears near the edge of the void structure.

\begin{figure}[htbp]
 \begin{center}
 \includegraphics[width=8cm,clip]{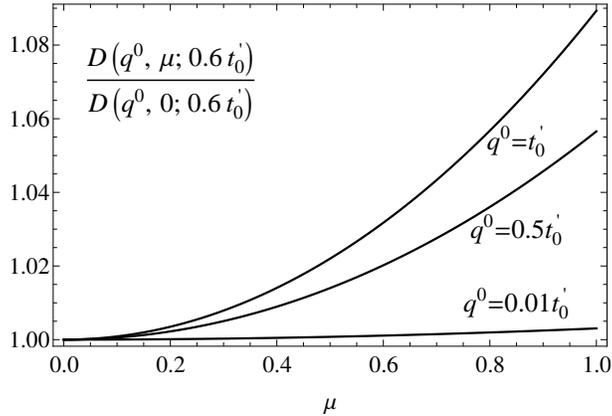}
 \end{center}
 \caption{
 The growth factor $D(q^0,\mu;r_{\rm o}'=0.6t_0')$ normalized 
 by its value of $\mu=0$ 
 at the three moments $q^0=0.01t_0'$, $q^0=0.5t_0'$ and $q^0=t_0'$ 
 is depicted as functions of $\mu$, in the case of the toy background model.
 }
 \label{5:fig_result1}
\end{figure}
\begin{figure}[htbp]
 \begin{center}
 \includegraphics[width=8cm,clip]{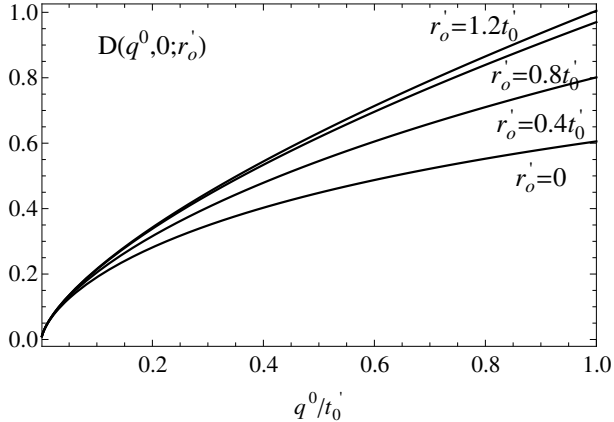}
 \end{center}
 \caption{
 The growth factors, $D(q^0,\mu=0;r_{\rm o}')$ as functions of $q^0/t_0'$ at 
the four radial positions specified by $r_{\rm o}'=0$, $r_{\rm o}'=0.4t_0'$, $r_{\rm o}'=0.8t_0'$ 
and $r_{\rm o}=1.2t_0'$, respectively, in the case of the toy background model .
 }
 \label{5:fig_result2}
\end{figure}

By numerically solving Eqs.~\eqref{5:sca1}--\eqref{5:sca3} 
with the initial conditions~\eqref{5:initial1}--\eqref{5:initial3} for the toy background model,
we obtained the growth factor $D(q^0,\mu;r_{\rm o}')$ in the toy background model.

In Fig.~\ref{5:fig_result1}, we plot the growth factor 
normalized by its value of $\mu=0$ with $r_{\rm o}'=0.6t_0'$
at the three moments, $q^0=0.01t_0'$, $q^0=0.5t_0'$ and $q^0=t_0'$, respectively, as functions of $\mu$. 
From this figure, we see that the anisotropy of the growth factor, 
that is, its $\mu$ dependence, grows as time goes on.
At the present time $q^0=t_0'$, the anisotropy of the growth factor is about 10\%.
Thus, we may conclude that non-negligible effects of the background 
anisotropy on the growth factor appear in the toy background model and 
maybe also in the typical huge void universe model. 
It is seen from Fig.~\ref{5:fig_result1} that the growth factor with $\mu=1$ is larger that that with $\mu=0$.
This fact implies that the growth rate of density contrast in the radial direction 
is larger than that in the transverse direction.
This is because the expansion rate of the radial direction, $H_\parallel$, 
is smaller than that of the transverse direction, $H_\bot$, in the toy background model
(see Fig.~\ref{5:fig_TOY3}).

In Fig.~\ref{5:fig_result2}, we depict the growth factor of $\mu=0$ 
as functions of $q^0/t_0'$ at the four radial positions specified by 
$r_{\rm o}'=0$, $r_{\rm o}'=0.4t_0'$, $r_{\rm o}'=0.8t_0'$ and $r_{\rm o}'=1.2t_0'$.
It is seen from this figure that the growth rate is an increasing function of the radial coordinate $r'$.
This behavior is understood as a consequence of the fact that the background energy density $\bar{\rho}$ 
is a monotonically increasing function of $r'$ in the toy background model, 
since the growth rate of perturbations are monotonically 
increasing function of the density parameter $\Omega_{\rm M}$ in the case of the dust-filled FLRW model.
It should be noted that the growth factor $D(q^0,0;r_{\rm o}')$ at the radial position $r_{\rm o}'=1.2t_0'$
coincides with that in the Einstein-de Sitter model.

\section{Conclusion and Discussion}\label{sec5}

We have studied the Newtonian perturbations in the huge void universe model 
based on the LTB solution.   
First, we introduced the Fermi-normal coordinates in which 
all physical and geometrical quantities are expressed in the form of the Maclaurin series. 
In this coordinate system, the effects of the background spacetime curvature   
on the perturbations are treated in the perturbative manner 
as long as the length scale of the perturbations of our interest is much smaller 
than the spacetime curvature radius of the background spacetime. 
This approximation scheme is called the tidal approximation. 
By the tidal approximation, we have shown that
the effects of the spacetime curvature of the background huge void can be  
significantly different from those in the homogeneous and isotropic universe model.
This results imply that the local FLRW approximation~\cite{Zibin:2008vj,Dunsby:2010ts}
in which the geometry of a sufficiently small region is 
assumed to be the same as that of the FLRW universe 
is not necessarily applicable to the huge void universe model.

The definition of the Newtonian perturbation was given in \S~\ref{sec1}; its self-gravity is 
so weak that the linear approximation is applicable to the metric perturbations; the relative 
velocities between the elements of the perturbations are much smaller than the speed of light; 
its size is much smaller than the spacetime curvature radius of the background. The last condition 
guarantees the applicability of the tidal approximation to the Newtonian perturbations. 
As in the case of the cosmological Newtonian approximation, the Newtonian perturbations in the 
huge void universe model are characterized by two small parameters denoted by 
$\epsilon$ and $\kappa$, respectively, whose definitions are given by Eqs.~\eqref{5:twopara-e} 
and \eqref{5:twopara-k}. 
Then, by using $\epsilon$ and $\kappa$ as expansion parameters of the perturbative treatment,  
we derived the basic equations for the Newtonian perturbations up to the leading order: 
the conservation law of the energy, the Euler equations for the weakly self-gravitating dust and 
the Poisson equation for the gravitational potential. 
The effects of the background tidal forces appear in the derived basic equations in the form of the 
anisotropic volume expansion of the background universe model. 
These results imply that the evolution of Newtonian perturbations 
in the huge void universe model can significantly differ from that in the homogeneous 
and isotropic universe model.

We studied the behavior of the density contrast whose amplitude   
is so small that the linear approximation is available. It is worthwhile to notice 
that this subject is very non-trivial, since the analysis of the linear perturbations 
in the LTB solution is, in general, not so easy.  
By adopting the spatial derivatives of the 3-velocity field instead of the 3-velocity field itself, and 
changing the coordinates of the background to the Lagrangian coordinates, 
the basic equations for the Newtonian perturbations are rewritten in the form of    
Eqs.~\eqref{5:L1a}--\eqref{5:L1e} which form a set of the partial differential equations 
with respect to the Lagrangian coordinates. A remarkable feature is that the coefficients 
of perturbation variables 
in these differential equations depend only on the Lagrangian time. By virtue of this 
feature, we can solve this set of the ordinary differential equations by performing the 
Fourier transformation with respect to the Lagrangian spatial coordinates, since 
each Fourier mode is decoupled from the other Fourier modes. 
Equations~\eqref{5:L2a}--\eqref{5:L2e} obtained by 
the Fourier transformation of Eqs.~\eqref{5:L1a}--\eqref{5:L1e} form a set of the 
ordinary differential equations with respect to the Lagrangian time. 
By contrast to the basic equations for the perturbations in the LTB solution 
without any other approximations except for the linearization, 
the basic equations obtained by the approximation scheme 
developed here are very simple and tractable.  
Then we solved the set of the basic equations \eqref{5:L2a}--\eqref{5:L2e} numerically 
and revealed the evolution of the vorticity field and the density contrast. 
We have shown that the vorticity field necessarily 
decays by the cosmic volume expansion in the case 
of the huge void universe model. 
This means that the vector mode of the Newtonian perturbations contains the decaying mode only 
in the case of the present background model.

We have shown in Figs.~\ref{5:fig_result1} and \ref{5:fig_result2} 
that the growth factor of density contrast significantly depends on the direction of the wave vector, 
$\mu$, and the radial position of the perturbations, $r_{\rm o}'$. 
The $\mu$- and $r_{\rm o}'$-dependences reflect 
the anisotropic volume expansion and the inhomogeneities 
in the radial direction of the background huge void, respectively. 
These properties of the evolution of density contrast are consistent with 
the results obtained in our previous works~\cite{Nishikawa:2012we,Nishikawa:2013rna} 
in which a different approximation scheme is adopted. 
Since the growth factor of the density contrast in the homogeneous and isotropic universe model 
is a function of only $t$, the dependence of the growth factor on $\mu$ and $r_{\rm o}'$ 
in the huge void universe model can be a strong discriminator
for these two models.

\begin{figure}[htbp]
 \begin{center}
 \includegraphics[width=7cm,clip]{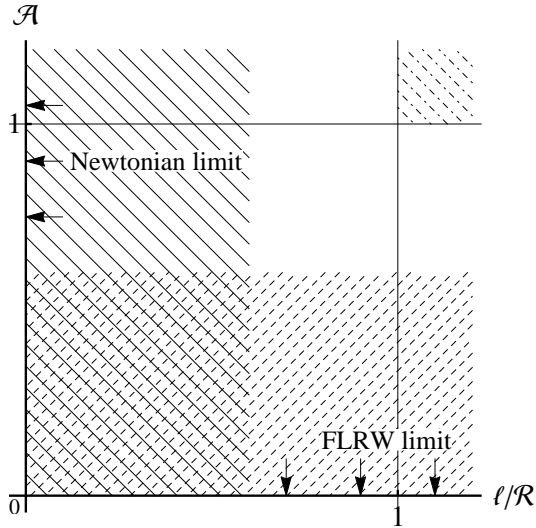}
 \end{center}
 \caption{A schematic picture representing a relation between the analysis for perturbations
 developed in our previous paper and that in this paper,
 where $\mathcal{A}$, $\ell$ and $\mathcal{R}$ denote the amplitude of inhomogeneity of 
 the background huge void model, the length scale of perturbations and the curvature radius of the 
 background .
 The region shaded by dashed lines 
 and the region shaded by solid lines represent domains in which the approximation schemes 
 adopted in our previous work and in this paper, respectively.
 }
 \label{6:fig}
\end{figure}

In our previous works~\cite{Nishikawa:2012we,Nishikawa:2013rna}, we studied 
the linear perturbations in the huge void universe model without any distinctions between relativistic and 
non-relativistic perturbations, by treating the background huge void as an isotropic perturbation 
in the background model of the homogeneous and isotropic universe filled with dust. 
By contrast, the method of analysis based on the tidal approximation is applicable, even if   
the background void structure is highly nonlinear.
In Fig.~\ref{6:fig}, we have shown a schematic picture representing the relation between our 
two independent approaches. Here, $\mathcal{A}$ denotes the amplitude of radial inhomogeneities
in the background huge void universe model, that is, the huge void itself, and $\ell/\mathcal{R}$ 
denotes the ratio of the length scale of perturbations to the 
spacetime curvature radius of the background which is assumed to be 
almost the same as the size of the void  
in the case of the huge void universe model. 
The region shaded by dashed lines in Fig.~\ref{6:fig} represents the domain 
in which the approximation scheme adopted in our previous 
works \cite{Nishikawa:2012we,Nishikawa:2013rna} is applicable.  
The region shaded by solid lines in Fig.~\ref{6:fig} represents 
the domain in which the approximation scheme adopted in the present paper 
is applicable. 
We can see that the region shaded by dot-dashed lines in Fig.~\ref{6:fig} is the domain to which 
neither the approximation scheme adopted in our previous works nor that adopted in the 
present paper is applicable. 
The system included in this region is composed of relativistic perturbations in the highly 
non-linear huge void. The study of perturbations in this region  
seems to be difficult without invoking the numerical simulations, 
and we leave it for a future work.

\section*{Acknowledgments}
K.N was supported in part by JSPS Grant-in-Aid for Scientifc Research (C) (No. 25400265).




\begin{thebibliography}{99}


\bibitem{Bull:2012zx} 
  P.~Bull and T.~Clifton,
  Phys.\ Rev.\ D {\bf 85}, 103512 (2012).


\bibitem{Celerier:1999hp}
  M.~N.~Celerier,
  Astron.\ Astrophys.\  {\bf 353}, 63 (2000).


\bibitem{Celerier:2009sv}
  M.~N.~Celerier, K.~Bolejko and A.~Krasinski,
  Astron.\ Astrophys.\  {\bf 518}, A21 (2010).


\bibitem{Clifton:2008hv}
  T.~Clifton, P.~G.~Ferreira and K.~Land,
  Phys.\ Rev.\ Lett.\  {\bf 101}, 131302 (2008).


\bibitem{Goodwin:1999ej}
  S.~P.~Goodwin, P.~A.~Thomas, A.~J.~Barber, J.~Gribbin and L.~I.~Onuora,
  arXiv:astro-ph/9906187.


\bibitem{Iguchi:2001sq}
  H.~Iguchi, T.~Nakamura and K.~i.~Nakao,
  Prog.\ Theor.\ Phys.\  {\bf 108}, 809 (2002).


\bibitem{Kolb:2009hn}
  E.~W.~Kolb and C.~R.~Lamb,
  arXiv:0911.3852 [astro-ph.CO].


\bibitem{Mustapha:1998jb}
  N.~Mustapha, C.~Hellaby, G.~F.~R.~Ellis,
  Mon.\ Not.\ Roy.\ Astron.\ Soc.\  {\bf 292}, 817-830 (1997).


\bibitem{Tomita:1999qn}
  K.~Tomita,
  Astrophys.\ J.\  {\bf 529}, 38 (2000).


\bibitem{Tomita:2000jj}
  K.~Tomita,
  Mon.\ Not.\ Roy.\ Astron.\ Soc.\  {\bf 326}, 287 (2001).


\bibitem{Tomita:2001gh}
  K.~Tomita,
  Prog.\ Theor.\ Phys.\  {\bf 106}, 929 (2001).


\bibitem{Vanderveld:2006rb}
  R.~A.~Vanderveld, E.~E.~Flanagan and I.~Wasserman,
  Phys.\ Rev.\  D {\bf 74}, 023506 (2006).


\bibitem{Yoo:2008su}
  C.~M.~Yoo, T.~Kai and K.~i.~Nakao,
  Prog.\ Theor.\ Phys.\  {\bf 120}, 937 (2008).


\bibitem{Yoo:2010qn}
  C.~-M.~Yoo,
  Prog.\ Theor.\ Phys.\  {\bf 124}, 645-665 (2010).


\bibitem{Alexander:2007xx}
  S.~Alexander, T.~Biswas, A.~Notari and D.~Vaid,
  JCAP {\bf 0909}, 025 (2009).


\bibitem{Alnes:2005rw}
  H.~Alnes, M.~Amarzguioui and O.~Gron,
  Phys.\ Rev.\  D {\bf 73}, 083519 (2006).


\bibitem{Biswas:2010xm}
  T.~Biswas, A.~Notari and W.~Valkenburg,
  JCAP {\bf 1011}, 030 (2010).


\bibitem{Bolejko:2008cm} 
  K.~Bolejko and J.~S.~B.~Wyithe,
  JCAP {\bf 0902}, 020 (2009).


\bibitem{Clarkson:2010ej}
  C.~Clarkson and M.~Regis,
  JCAP {\bf 1102}, 013 (2011).


\bibitem{GarciaBellido:2008nz}
  J.~Garcia-Bellido and T.~Haugboelle,
  JCAP {\bf 0804}, 003 (2008).


\bibitem{Marra:2011ct}
  V.~Marra and A.~Notari,
  Class.\ Quant.\ Grav.\  {\bf 28}, 164004 (2011).


\bibitem{Marra:2010pg}
  V.~Marra and M.~Paakkonen,
  JCAP {\bf 1012}, 021 (2010).


\bibitem{Moss:2010jx}
  A.~Moss, J.~P.~Zibin and D.~Scott,
  Phys.\ Rev.\  D {\bf 83}, 103515 (2011).


\bibitem{Nadathur:2010zm} 
  S.~Nadathur and S.~Sarkar,
  Phys.\ Rev.\ D {\bf 83}, 063506 (2011).


\bibitem{Yoo:2010qy}
  C.~M.~Yoo, K.~i.~Nakao and M.~Sasaki,
  JCAP {\bf 1007}, 012 (2010).


\bibitem{Zibin:2008vk}
  J.~P.~Zibin, A.~Moss and D.~Scott,
  Phys.\ Rev.\ Lett.\  {\bf 101}, 251303 (2008).


\bibitem{GarciaBellido:2008yq}
  J.~Garcia-Bellido and T.~Haugboelle,
  JCAP {\bf 0909}, 028 (2009).


\bibitem{Zumalacarregui:2012pq} 
  M.~Zumalacarregui, J.~Garcia-Bellido and P.~Ruiz-Lapuente,
  JCAP {\bf 1210}, 009 (2012).


\bibitem{Bull:2011wi}
  P.~Bull, T.~Clifton and P.~G.~Ferreira,
  Phys.\ Rev.\ D {\bf 85}, 024002 (2012).


\bibitem{GarciaBellido:2008gd}
  J.~Garcia-Bellido and T.~Haugboelle,
  JCAP {\bf 0809}, 016 (2008).


\bibitem{Moss:2011ze}
  A.~Moss and J.~P.~Zibin,
  Class.\ Quant.\ Grav.\  {\bf 28}, 164005 (2011).


\bibitem{Yoo:2010ad}
  C.~M.~Yoo, K.~i.~Nakao and M.~Sasaki,
  JCAP {\bf 1010}, 011 (2010).


\bibitem{Zhang:2010fa}
  P.~Zhang and A.~Stebbins,
  Phys.\ Rev.\ Lett.\  {\bf 107}, 041301 (2011).


\bibitem{Ade:2013opi} 
  P.~A.~R.~Ade {\it et al.}  [Planck Collaboration],
  Astron.\ Astrophys.\  {\bf 561}, A97 (2014).


\bibitem{Adachi:2011vu} 
  M.~Adachi and M.~Kasai,
  Prog.\ Theor.\ Phys.\  {\bf 127}, 145 (2012).


\bibitem{Alnes:2006uk}
  H.~Alnes and M.~Amarzguioui,
  Phys.\ Rev.\  D {\bf 75}, 023506 (2007).


\bibitem{Bolejko:2011jc} 
  K.~Bolejko, M.~-N.~Celerier and A.~Krasinski,
  Class.\ Quant.\ Grav.\  {\bf 28}, 164002 (2011).


\bibitem{Bolejko:2005fp}
  K.~Bolejko,
  PMC Phys.\  A {\bf 2}, 1 (2008).


\bibitem{Caldwell:2013fua} 
  R.~R.~Caldwell and N.~A.~Maksimova,
  Phys.\ Rev.\ D {\bf 88}, no. 10, 103502 (2013).


\bibitem{Celerier:2012xr} 
  M.~-N.~Celerier,
  arXiv:1203.2814 [astro-ph.CO].


\bibitem{Clarkson:2012bg} 
  C.~Clarkson,
  Comptes Rendus Physique {\bf 13}, 682 (2012).


\bibitem{dePutter:2012zx} 
  R.~de Putter, L.~Verde and R.~Jimenez,
  JCAP {\bf 1302}, 047 (2013).


\bibitem{Dunsby:2010ts}
  P.~Dunsby, N.~Goheer, B.~Osano and J.~P.~Uzan,
  JCAP {\bf 1006}, 017 (2010).


\bibitem{Enqvist:2009hn}
  K.~Enqvist, M.~Mattsson and G.~Rigopoulos,
  JCAP {\bf 0909}, 022 (2009).


\bibitem{Enqvist:2006cg}
  K.~Enqvist and T.~Mattsson,
  JCAP {\bf 0702}, 019 (2007).


\bibitem{Goto:2011ru}
  H.~Goto and H.~Kodama,
  Prog.\ Theor.\ Phys.\  {\bf 125}, 815 (2011).


\bibitem{Heavens:2011mr} 
  A.~F.~Heavens, R.~Jimenez and R.~Maartens,
  JCAP {\bf 1109}, 035 (2011).


\bibitem{Quartin:2009xr}
  M.~Quartin and L.~Amendola,
  Phys.\ Rev.\  D {\bf 81}, 043522 (2010).


\bibitem{Regis:2010iq}
  M.~Regis and C.~Clarkson,
  Gen. Rel. Grav. {\bf 44}, 567 (2012).


\bibitem{Romano:2009mr}
  A.~E.~Romano,
  Phys.\ Rev.\  D {\bf 82}, 123528 (2010).


\bibitem{Romano:2010nc} 
  A.~E.~Romano, M.~Sasaki and A.~A.~Starobinsky,
  Eur. Phys. J. {\bf C72}, 2242 (2012).


\bibitem{Romano:2011mx} 
  A.~E.~Romano and P.~Chen,
  JCAP {\bf 1110}, 016 (2011).


\bibitem{Tanimoto:2009mz}
  M.~Tanimoto, Y.~Nambu and K.~Iwata,
  arXiv:0906.4857 [astro-ph.CO].


\bibitem{Tomita:2009ar} 
  K.~Tomita,
  arXiv:0906.1325 [astro-ph.CO].


\bibitem{Yagi:2012vx} 
  K.~Yagi, A.~Nishizawa and C.~-M.~Yoo,
  J.\ Phys.\ Conf.\ Ser.\  {\bf 363}, 012056 (2012).


\bibitem{Zibin:2011ma}
  J.~P.~Zibin,
  Phys.\ Rev.\  D {\bf 84}, 123508 (2011).



\bibitem{Valkenburg:2013qwa} 
  W.~Valkenburg, M.~Kunz and V.~Marra,
  Phys.\ Dark Univ.\  {\bf 2}, 219 (2013).


\bibitem{Valkenburg:2012td} 
  W.~Valkenburg, V.~Marra and C.~Clarkson,
  MNRAS {\bf 438}, (2014) L6.



\bibitem{Alonso:2012ds} 
  D.~Alonso, J.~Garcia-Bellido, T.~Haugboelle and A.~Knebe,
  Phys.\ Dark Univ.\ {\bf 1}, 24 (2012).


\bibitem{Alonso:2010zv}
  D.~Alonso, J.~Garcia-Bellido, T.~Haugbolle and J.~Vicente,
  Phys.\ Rev.\  D {\bf 82}, 123530 (2010).


\bibitem{Clarkson:2009sc}
  C.~Clarkson, T.~Clifton and S.~February,
  JCAP {\bf 0906}, 025 (2009).


\bibitem{February:2012fp} 
  S.~February, C.~Clarkson and R.~Maartens,
  JCAP {\bf 1303}, 023 (2013).


\bibitem{February:2013qza} 
  S.~February, J.~Larena, C.~Clarkson and D.~Pollney,
  arXiv:1311.5241 [astro-ph.CO].


\bibitem{Zibin:2008vj}
  J.~P.~Zibin,
  Phys.\ Rev.\  D {\bf 78}, 043504 (2008).



\bibitem{Nishikawa:2012we} 
  R.~Nishikawa, C.~-M.~Yoo and K.~-i.~Nakao,
  Phys.\ Rev.\ D {\bf 85}, 103511 (2012).


\bibitem{Nishikawa:2013rna} 
  R.~Nishikawa, C.~-M.~Yoo and K.~-i.~Nakao,
  Phys.\ Rev.\ D {\bf 88}, 123520 (2013).


\bibitem{Nishikawa:2014sga} 
  R.~Nishikawa, K.~i.~Nakao and C.~M.~Yoo,
  arXiv:1407.4899 [astro-ph.CO].




\bibitem{peebles}
  P.~J.~E.~Peebles,
  ``The Large-Scale Structure of the Universe,''
  (Princeton University Press, Princeton, 1980).

\bibitem{Futamase:1989} 
  T.~Futamase,
  Mon.\ Not.\ Roy.\ Astron.\ Soc.\  {\bf 237}, 187 (1989).


\bibitem{Hwang:2005mg} 
  J.~-C.~Hwang, H.~Noh and D.~Puetzfeld,
  JCAP {\bf 0803}, 010 (2008).


\bibitem{Shibata:1995dg} 
  M.~Shibata and H.~Asada,
  Prog.\ Theor.\ Phys.\  {\bf 94}, 11 (1995).


\bibitem{Tomita:1988} 
  K.~Tomita,
  Prog.\ Theor.\ Phys.\  {\bf 79}, 2 (1988).






\bibitem{MM:1963}
F.K.~Manasse and C.W.~Misner,
J.\ Math.\ Phys.\ {\bf 4}, 735 (1963). 

\bibitem{Ishii:2005xq} 
  M.~Ishii, M.~Shibata and Y.~Mino,
  Phys.\ Rev.\ D {\bf 71}, 044017 (2005).




\bibitem{Baldauf:2011bh} 
  T.~Baldauf, U.~Seljak, L.~Senatore and M.~Zaldarriaga,
  JCAP {\bf 1110}, 031 (2011).


\bibitem{Klein:2007xj} 
  D.~Klein and P.~Collas,
  Class.\ Quant.\ Grav.\  {\bf 25}, 145019 (2008).


\bibitem{toolkit}
  E.~Poisson,
  ``A relativist's toolkit,''
  (Cambridge University Press, Cambridge, 2004).


\bibitem{Schmidt:2012nw} 
  F.~Schmidt and D.~Jeong,
  Phys.\ Rev.\ D {\bf 86}, 083513 (2012).


\bibitem{Mashhoon:2007qm} 
  B.~Mashhoon, N.~Mobed and D.~Singh,
  Class.\ Quant.\ Grav.\  {\bf 24}, 5031 (2007).



\bibitem{wald}
  R.~M.~Wald,
  ``General Relativity,''
  (The University of Chicago Press, Chicago, 1984).


\bibitem{Gerlach:1980}
  U.H.~Gerlach and U.K.~Sengupta,
  Phys.\ Rev.\  D {\bf 19}, 2268 (1979).





\end{thebibliography}
\end{document}